\documentclass[preprint, superscriptaddress, amsmath,amssymb]{revtex4}

\usepackage{booktabs}
\usepackage{color,graphicx}
\usepackage{amsmath}
\usepackage{dcolumn}
\usepackage{CJK}                 
\usepackage{bm}                  
\usepackage{soul}
\usepackage{enumerate}
\usepackage{multirow}
\usepackage{mathrsfs}
\usepackage{braket}
\usepackage{longtable}
\usepackage{pdflscape}

\usepackage[pdfstartview=FitH,
            CJKbookmarks=true,
            bookmarksnumbered=true,
            bookmarksopen=true,
            colorlinks,
            pdfborder=001,
            linkcolor=blue,
            anchorcolor=blue,
            citecolor=blue
            ]{hyperref}

\usepackage{float}

\newcommand{\cred}{\color{red}}
\newcommand{\fra}[1]{\vspace*{0.05in} \hrule \vspace*{0.05in}
\noindent {\color{red}\textsf{F: #1}} \vspace*{0.05in}\hrule
\vspace*{0.05in}}
\newcommand{\fraadd}[1]{{\color{red} #1}}

\newcommand{\adr}[1]{\vspace*{0.05in} \hrule \vspace*{0.05in}
\noindent {\color{green}\textsf{A: #1}} \vspace*{0.05in}\hrule
\vspace*{0.05in}}
\newcommand{\adradd}[1]{{\color{green} #1}}
\newcommand{\adrst}[1]{\setstcolor{green} \st{#1}} 

\newcommand{\jer}[1]{\vspace*{0.05in} \hrule \vspace*{0.05in}
\noindent {\color{blue}\textsf{J: #1}} \vspace*{0.05in}\hrule
\vspace*{0.05in}} 
\newcommand{\jerhl}[1]{{\sethlcolor{blue} \hl{#1}} } 
\newcommand{\jeradd}[1]{{\color{blue} #1}} 
\newcommand{\jerst}[1]{\setstcolor{blue} \st{#1}} 

\begin{document}
\begin{CJK*}{GBK}{song}

\title{Nuclear mass table in deformed relativistic Hartree-Bogoliubov theory in continuum, II: Even-$Z$ nuclei}

\author{Peng Guo}
\affiliation{State Key Laboratory of Nuclear Physics and Technology, School of Physics, Peking University, Beijing 100871, China}

\author{Xiaojie Cao}
\affiliation{Department of Physics, School of Science, Tianjin University, Tianjin 300354, China}

\author{Kangmin Chen}
\affiliation{School of Physics and Mechatronics Engineering, Guizhou Minzu University, Guiyang 550025, China}

\author{Zhihui Chen}
\affiliation{Department of Physics, The University of Hong Kong, Pokfulam 999077, Hong Kong, China}

\author{Myung-Ki Cheoun}
\affiliation{Department of Physics and Origin of Matter and Evolution of Galaxy (OMEG) Institute, Soongsil University, Seoul 06978, Republic of Korea}

\author{Yong-Beom Choi}
\affiliation{Center for Innovative Physicist Education and Research, Extreme Physics Institute, and Department of Physics, Pusan National University, Busan 46241, Republic of Korea}

\author{Pak Chung Lam}
\affiliation{Department of Physics, The University of Hong Kong, Pokfulam 999077, Hong Kong, China}

\author{Wenmin Deng}
\affiliation{Department of Physics, Beijing Normal University, Beijing 100875, China}

\author{Jianmin Dong}
\affiliation{Institute of Modern Physics, Chinese Academy of Sciences, Lanzhou 730000, China}
\affiliation{School of Physics, University of Chinese Academy of Sciences, Beijing 100049, China}

\author{Pengxiang Du}
\affiliation{College of Physics, Jilin University, Changchun 130012, China}

\author{Xiaokai Du}
\affiliation{State Key Laboratory of Nuclear Physics and Technology, School of Physics, Peking University, Beijing 100871, China}

\author{Kangda Duan}
\affiliation{Institute of Modern Physics, Chinese Academy of Sciences, Lanzhou 730000, China}

\author{Xiaohua Fan}
\affiliation{School of Physical Science and Technology, Southwest University, Chongqing 400715, China}

\author{Wei Gao}
\affiliation{School of Physics and Microelectronics, Zhengzhou University, Zhengzhou 450001, China}

\author{Lisheng Geng}
\affiliation{School of Physics, Beihang University, Beijing 102206, China}
\affiliation{Peng Huanwu Collaborative Center for Research and Education, Beihang University, Beijing 100191, China}
\affiliation{Beijing Key Laboratory of Advanced Nuclear Materials and Physics, Beihang University, Beijing 102206, China}
\affiliation{Southern Center for Nuclear-Science Theory (SCNT), Institute of Modern Physics, Chinese Academy of Sciences, Huizhou 516000, China}

\author{Eunja Ha}
\affiliation{Department of Physics, Hanyang University, Seoul 04763, Republic of Korea}

\author{Xiao-Tao He}
\affiliation{College of Materials Science and Technology, Nanjing University of Aeronautics and Astronautics, Nanjing 210016, China}

\author{Jinniu Hu}
\affiliation{School of Physics, Nankai University, Tianjin 300071, China}

\author{Jingke Huang}
\affiliation{School of Physics and Microelectronics, Zhengzhou University, Zhengzhou 450001, China}

\author{Kun Huang}
\affiliation{College of Materials Science and Technology, Nanjing University of Aeronautics and Astronautics, Nanjing 210016, China}

\author{Yanan Huang}
\affiliation{School of Nuclear Science and Technology, Lanzhou University, Lanzhou 730000, China}

\author{Zidan Huang}
\affiliation{School of Physics and Microelectronics, Zhengzhou University, Zhengzhou 450001, China}

\author{Kim Da Hyung}
\affiliation{Department of Physics, The University of Hong Kong, Pokfulam 999077, Hong Kong, China}

\author{Hoi Yat Chan}
\affiliation{Department of Physics, The University of Hong Kong, Pokfulam 999077, Hong Kong, China}

\author{Xiaofei Jiang}
\affiliation{State Key Laboratory of Nuclear Physics and Technology, School of Physics, Peking University, Beijing 100871, China}

\author{Seonghyun Kim}
\affiliation{Department of Physics and Origin of Matter and Evolution of Galaxy Institute, Soongsil University, Seoul 06978, Republic of Korea}

\author{Youngman Kim}
\affiliation{Center for Exotic Nuclear Studies, Institute for Basic Science, Daejeon 34126, Republic of Korea}

\author{Chang-Hwan Lee}
\affiliation{Center for Innovative Physicist Education and Research, Extreme Physics Institute, and Department of Physics, Pusan National University, Busan 46241, Republic of Korea}

\author{Jenny Lee}
\affiliation{Department of Physics, The University of Hong Kong, Pokfulam 999077, Hong Kong, China}

\author{Jian Li}
\affiliation{College of Physics, Jilin University, Changchun 130012, China}

\author{Minglong Li}
\affiliation{Department of Physics, Graduate School of Science, The University of Tokyo, Tokyo 113-0033, Japan}

\author{Zhipan Li}
\affiliation{School of Physical Science and Technology, Southwest University, Chongqing 400715, China}

\author{Zhengzheng Li}
\affiliation{State Key Laboratory of Nuclear Physics and Technology, School of Physics, Peking University, Beijing 100871, China}

\author{Zhanjiang Lian}
\affiliation{China Institute of Atomic Energy, Beijing 102413, China}

\author{Haozhao Liang}
\affiliation{Department of Physics, Graduate School of Science, The University of Tokyo, Tokyo 113-0033, Japan}
\affiliation{RIKEN Interdisciplinary Theoretical and Mathematical Sciences Program (iTHEMS), Wako 351-0198, Japan}

\author{Lang Liu}
\affiliation{School of Science, Jiangnan University, Wuxi 214122, China}

\author{Xiao Lu}
\affiliation{CAS Key Laboratory of Theoretical Physics, Institute of Theoretical Physics, Chinese Academy of Sciences, Beijing 100190, China}

\author{Zhi-Rui Liu}
\affiliation{College of Materials Science and Technology, Nanjing University of Aeronautics and Astronautics, Nanjing 210016, China}

\author{Jie Meng} \email{mengj@pku.edu.cn}
\affiliation{State Key Laboratory of Nuclear Physics and Technology, School of Physics, Peking University, Beijing 100871, China}

\author{Ziyan Meng}
\affiliation{State Key Laboratory of Nuclear Physics and Technology, School of Physics, Peking University, Beijing 100871, China}

\author{Myeong-Hwan Mun}
\affiliation{Department of Physics and Origin of Matter and Evolution of Galaxy Institute, Soongsil University, Seoul 06978, Republic of Korea}

\author{Yifei Niu}
\affiliation{School of Nuclear Science and Technology, Lanzhou University, Lanzhou 730000, China}

\author{Zhongming Niu}
\affiliation{School of Physics and Optoelectronics Engineering, Anhui University, Hefei 230601, China}

\author{Cong Pan}
\affiliation{Department of Physics, Anhui Normal University, Wuhu 241000, China}
\affiliation{State Key Laboratory of Nuclear Physics and Technology, School of Physics, Peking University, Beijing 100871, China}

\author{Jing Peng}
\affiliation{Department of Physics, Beijing Normal University, Beijing 100875, China}

\author{Xiaoying Qu}
\affiliation{School of Physics and Mechatronics Engineering, Guizhou Minzu University, Guiyang 550025, China}

\author{Panagiota Papakonstantinou}
\affiliation{Institute for Rare Isotope Science, Institute for Basic Science, Daejeon 34000, Republic of Korea}

\author{Tianshuai Shang}
\affiliation{College of Physics, Jilin University, Changchun 130012, China}

\author{Xinle Shang}
\affiliation{Institute of Modern Physics, Chinese Academy of Sciences, Lanzhou 730000, China}
\affiliation{School of Physics, University of Chinese Academy of Sciences, Beijing 100049, China}

\author{Caiwan Shen}
\affiliation{School of Science, Huzhou University, Huzhou 313000, China}

\author{Guofang Shen}
\affiliation{School of Physics, Beihang University, Beijing 102206, China}

\author{Tingting Sun}
\affiliation{School of Physics and Microelectronics, Zhengzhou University, Zhengzhou 450001, China}

\author{Xiang-Xiang Sun}
\affiliation{School of Nuclear Science and Technology, University of Chinese Academy of Sciences, Beijing 100049, China}
\affiliation{CAS Key Laboratory of Theoretical Physics, Institute of Theoretical Physics, Chinese Academy of Sciences, Beijing 100190, China}

\author{Sibo Wang}
\affiliation{Department of Physics, Chongqing Key Laboratory for Strongly Coupled Physics, Chongqing University, Chongqing 401331, China}

\author{Tianyu Wang}
\affiliation{School of Mechatronics Engineering, Guizhou Minzu University, Guiyang 550025, China}

\author{Yiran Wang}
\affiliation{Department of Physics, School of Science, Tianjin University, Tianjin 300354, China}

\author{Yuanyuan Wang}
\affiliation{Mathematics and Physics Department, North China Electric Power University, Beijing 102206, China}

\author{Jiawei Wu}
\affiliation{College of Materials Science and Technology, Nanjing University of Aeronautics and Astronautics, Nanjing 210016, China}

\author{Liang Wu}
\affiliation{School of Physics and Microelectronics, Zhengzhou University, Zhengzhou 450001, China}

\author{Xinhui Wu}
\affiliation{Department of Physics, Fuzhou University, Fuzhou 350108, Fujian, China}

\author{Xuewei Xia}
\affiliation{School of Physics and Electronic Engineering, Center for Computational Sciences, Sichuan Normal University, Chengdu 610068, China}

\author{Huihui Xie}
\affiliation{College of Physics, Jilin University, Changchun 130012, China}

\author{Jiangming Yao}
\affiliation{School of Physics and Astronomy, Sun Yat-sen University, Zhuhai, 519082, China}
\affiliation{Facility for Rare Isotope Beams, Michigan State University, East Lansing MI, 48824-1321, USA}

\author{Kwan Yau Ip}
\affiliation{Department of Physics, The University of Hong Kong, Pokfulam 999077, Hong Kong, China}

\author{To Chung Yiu}
\affiliation{Department of Physics, The University of Hong Kong, Pokfulam 999077, Hong Kong, China}

\author{Jianghan Yu}
\affiliation{School of Science, Huzhou University, Huzhou 313000, China}

\author{Yangyang Yu}
\affiliation{School of Physics and Mechatronics Engineering, Guizhou Minzu University, Guiyang 550025, China}

\author{Kaiyuan Zhang}
\affiliation{Institute of Nuclear Physics and Chemistry, CAEP, Mianyang, Sichuan 621900, China}
\affiliation{State Key Laboratory of Nuclear Physics and Technology, School of Physics, Peking University, Beijing 100871, China}

\author{Shijie Zhang}
\affiliation{School of Physics and Microelectronics, Zhengzhou University, Zhengzhou 450001, China}

\author{Shuangquan Zhang}
\affiliation{State Key Laboratory of Nuclear Physics and Technology, School of Physics, Peking University, Beijing 100871, China}

\author{Wei Zhang}
\affiliation{School of Physics and Microelectronics, Zhengzhou University, Zhengzhou 450001, China}

\author{Xiaoyan Zhang}
\affiliation{School of Physics and Optoelectronics Engineering, Anhui University, Hefei 230601, China}

\author{Yanxin Zhang}
\affiliation{School of Physics and Astronomy, Sun Yat-sen University, Zhuhai, 519082, China}

\author{Ying Zhang}
\affiliation{Department of Physics, School of Science, Tianjin University, Tianjin 300354, China}

\author{Yingxun Zhang}
\affiliation{China Institute of Atomic Energy, Beijing 102413, China}
\affiliation{Guangxi Key Laboratory of Nuclear Physics and Technology, Guangxi Normal University, Guilin 541004, China}

\author{Zhenhua Zhang}
\affiliation{Mathematics and Physics Department, North China Electric Power University, Beijing 102206, China}

\author{Qiang Zhao}
\affiliation{Center for Exotic Nuclear Studies, Institute for Basic Science, Daejeon 34126, Republic of Korea}

\author{Yingchun Zhao}
\affiliation{State Key Laboratory of Nuclear Physics and Technology, School of Physics, Peking University, Beijing 100871, China}

\author{Ruyou Zheng}
\affiliation{School of Physics, Beihang University, Beijing 102206, China}

\author{Chang Zhou}
\affiliation{State Key Laboratory of Nuclear Physics and Technology, School of Physics, Peking University, Beijing 100871, China}

\author{Shan-Gui Zhou}
\affiliation{CAS Key Laboratory of Theoretical Physics, Institute of Theoretical Physics, Chinese Academy of Sciences, Beijing 100190, China}
\affiliation{School of Physical Sciences, University of Chinese Academy of Sciences, Beijing 100049, China}
\affiliation{School of Nuclear Science and Technology, University of Chinese Academy of Sciences, Beijing 100049, China}
\affiliation{Peng Huanwu Collaborative Center for Research and Education, Beihang University, Beijing 100191, China}

\author{Lianjian Zou}
\affiliation{School of Physics, Nankai University, Tianjin 300071, China}

\collaboration{DRHBc Mass Table Collaboration}

\begin{abstract}
The mass table in the deformed relativistic Hartree-Bogoliubov theory in continuum (DRHBc) with the PC-PK1 density functional has been established for even-$Z$ nuclei with $8\le Z\le120$, extended from the previous work for even-even nuclei~[Zhang $et~al.$ (DRHBc Mass Table Collaboration),~\href{https://www.sciencedirect.com/science/article/pii/S0092640X22000018}{At. Data Nucl. Data Tables 144, 101488 (2022)}].
The calculated binding energies, two-nucleon and one-neutron separation energies, root-mean-square (rms) radii of neutron, proton, matter, and charge distributions, quadrupole deformations, and neutron and proton Fermi surfaces are tabulated and compared with available experimental data. A total of 4829 even-$Z$ nuclei are predicted to be bound, with an rms deviation of 1.433 MeV from the 1244 mass data. Good agreement with the available experimental odd-even mass differences, $\alpha$ decay energies, and charge radii is also achieved. The description accuracy for nuclear masses and nucleon separation energies as well as the prediction for drip lines is compared with the results obtained from other relativistic and nonrelativistic density functional. The comparison shows that the DRHBc theory with PC-PK1 provides an excellent microscopic description for the masses of even-$Z$ nuclei. The systematics of the nucleon separation energies, odd-even mass differences, pairing energies, two-nucleon gaps, $\alpha$ decay energies, rms radii, quadrupole deformations, potential energy curves, neutron density distributions, and neutron mean-field potentials are discussed. 
\end{abstract}

\date{\today}

\maketitle
\tableofcontents


\section{Introduction}
In a previous paper~\cite{Zhang2022mass}, following the strategy and techniques presented in Ref.~\cite{Zhang2020PRC}, the nuclear mass table calculated by the deformed relativistic Hartree-Bogoliubov theory in continuum (DRHBc)~\cite{Zhou2010PRC,Li2012PRC,Li2012CPL} based on the density functional PC-PK1~\cite{Zhao2010PRC} was constructed for even-even nuclei.
A total of 2583 even-even nuclei with $8\le Z\le120$ have been predicted to be bound, and the root-mean-square deviation (rms) from experimental data is 1.518 MeV. 
With the DRHBc theory extended for odd nuclei in Ref.~\cite{Pan2022PRC}, this paper will report the latest results of the DRHBc calculations for even-$Z$ odd-$N$ nuclei with $8 \le Z \le 120$ and construct the DRHBc mass table for even-$Z$ nuclei.

The DRHBc theory, based on meson-exchange~\cite{Zhou2010PRC,Li2012PRC,Li2012CPL,Chen2012PRC} or point-coupling~\cite{Zhang2020PRC,Pan2022PRC} density functionals, treats the deformation degrees of freedom, pairing correlations, and continuum effects properly, and provides a microscopic description for both exotic and stable nuclei.
It has been applied to study the halos in carbon isotopes~\cite{SUN2018PLB,SUN2020NPA}, the connection between halo phenomena and nucleons in the classically forbidden region~\cite{ZhangK2019PRC}, the deformation effects on the location of neutron drip line from O to Ca isotopes~\cite{EJ2021IJMPE}, and the influence of nuclear rotation correction on $\alpha$-decay half-lives of superheavy nuclei~\cite{Theeb2024NPA}.
It predicts the bubble structure and shape coexistence in the $72\le Z \le 82$ region~\cite{Choi2022PRC,Kim2022PRC}, and peninsulas of stability beyond the two-neutron drip line~\cite{Zhang2021PRC,Pan2021PRC,He2021CPC}. 
Based on the DRHBc theory, the angular momentum projection has been implemented to investigate the low-lying spectra of neutron-rich magnesium isotopes~\cite{SunZhou2021PRC,Sun2021SB}, and the finite amplitude method has been developed to study the isoscalar giant monopole resonance in exotic nuclei~\cite{Sun2022PRC}. The optimization of the DRHBc theory has been studied for the multipole expansion of nuclear densities~\cite{Pan2019IJMPE}, the Dirac Woods-Saxon (DWS) basis optimized~\cite{ZaP2022PRC}, and the dynamical correlation energy and the nuclear shape evolution explored by the two-dimensional collective Hamiltonian (2DCH)~\cite{Sun2022CPC,ZhangXY2023PRC}.

The DRHBc theory for odd nuclei treats properly the blocking effect of the unpaired nucleon(s) with the orbital-fixed~\cite{Li2012CPL} or automatic blocking procedure~\cite{Pan2022PRC}.
It successfully describes halo nuclei $^{17,19}$B~\cite{Yang2021PRL,Sun2021PRC} and $^{37}$Mg~\cite{Zhang2023PLB} self-consistently, the prolate shape dominance in Te, Xe, and Ba isotopes~\cite{Guo2023PRC}, and the odd-even shape staggering and kink structure of charge radii of Hg isotopes~\cite{Mun2023PLB}.
It predicts a possible two-neutron halo and the collapse of the $N = 28$ shell closure in $^{39}$Na~\cite{Zhang2023PRC}.
Based on the DRHBc theory for odd nuclei, the one-proton emissions in $^{148-151}$Lu have been studied with the WKB method~\cite{Xiao2023PLB}, and the deformed halo nuclei $^{31}$Ne~\cite{Zhong2022SCP} and $^{37}$Mg~\cite{An2024PLB} and the charge-changing cross sections of several $p$-shell nuclei on a carbon target have been investigated by the Glauber model~\cite{Zhao2023PLB}.
These successes greatly encourage the construction of the DRHBc mass table for odd nuclei.

In this paper, we present and discuss the DRHBc mass table for even-$Z$ nuclei with $8\le Z\le120$ including both even-$N$ and odd-$N$.
The DRHBc theoretical framework is briefly introduced in Sec.~\ref{theory}.
Numerical details are listed in Sec.~\ref{numerical}. Extensive results and discussions are compiled in Sec.~\ref{results}, including nuclear masses, nucleon separation energies, odd-even mass differences, rms radii, quadrupole deformations, potential energy curves, neutron density distributions, and neutron potentials.
Finally, a summary is given in Sec.~\ref{summary}.


\section{Theoretical framework} \label{theory}

The details of the DRHBc theory with meson-exchange and point-coupling density functionals can be found in Refs.~\cite{Li2012PRC} and \cite{Zhang2020PRC,Pan2022PRC}, respectively. In the following, the theoretical framework of DRHBc will be introduced briefly.

The relativistic density functional theory starts from an effective Lagrangian with either the meson-exchange or point-coupling interaction~\cite{Meng2016RDFNS}. For the point-coupling interaction adopted, the Lagrangian reads
\begin{align}\label{Lagrangian}
\mathscr{L} = & \bar{\psi}(i\gamma_\mu\partial^\mu-m)\psi - \frac{1}{2}\alpha_S(\bar{\psi}\psi)(\bar{\psi}\psi)\notag \\
& -\frac{1}{2}\alpha_V(\bar{\psi}\gamma_\mu\psi)(\bar{\psi}\gamma^\mu\psi)-\frac{1}{2}\alpha_{TV}(\bar{\psi}\vec{\tau}\gamma_\mu\psi)(\bar{\psi}\vec{\tau}\gamma^\mu\psi)\notag \\
& -\frac{1}{3}\beta_S(\bar{\psi}\psi)^3-\frac{1}{4}\gamma_S(\bar{\psi}\psi)^4-\frac{1}{4}\gamma_V[(\bar{\psi}\gamma_\mu\psi)(\bar{\psi}\gamma^\mu\psi)]^2\notag \\
&-\frac{1}{2}\delta_S\partial_\nu(\bar{\psi}\psi)\partial^\nu(\bar{\psi}\psi)-\frac{1}{2}\delta_V\partial_\nu(\bar{\psi}\gamma_\mu\psi)\partial^\nu(\bar{\psi}\gamma^\mu\psi)\notag \\
&-\frac{1}{2}\delta_{TV}\partial_\nu(\bar{\psi}\vec{\tau}\gamma_\mu\psi)\partial^\nu(\bar{\psi}\vec{\tau}\gamma^\mu\psi)\notag \\
&-\frac{1}{4}F^{\mu\nu}F_{\mu\nu}-e\bar{\psi}\gamma^{\mu}\frac{1-\tau_3}{2} {\psi} A_{\mu},
\end{align}
where $M$ is the nucleon mass, $e$ is the charge unit, and $A_{\mu}$ and $F_{\mu \nu}$ are the four-vector potential and field strength tensor of the electromagnetic field, respectively. Here $\alpha_S$, $\alpha_V$, and $\alpha_{TV}$ represent the coupling constants for four-fermion terms, $\beta_S$, $\gamma_S$, and $\gamma_V$ are those for the high-order terms, and $\delta_S$, $\delta_V$, and $\delta_{TV}$ refer to those for gradient terms. The subscripts $S$, $V$, and $TV$ stand for scalar, vector, and isovector, respectively. The isovector-scalar channel is neglected since the inclusion of isovector-scalar does not improve the nuclear ground-state properties~\cite{TB2002PRC}.

From the Lagrangian density of Eq.~(\ref{Lagrangian}), the density functional can be constructed under the mean-field and no-sea approximation. 
By minimizing the density functional with respect to the densities, the Dirac equation for nucleons can be obtained within the relativistic mean-field framework.
The relativistic Hartree-Bogoliubov (RHB) equation~\cite{Kucharek1991ZPA} treats self-consistently the mean field and pairing correlations,  
\begin{equation}\label{RHB}
\left(\begin{matrix}
h_D-\lambda_\tau & \Delta \\
-\Delta^* &-h_D^*+\lambda_\tau
\end{matrix}\right)\left(\begin{matrix}
U_k\\
V_k
\end{matrix}\right)=E_k\left(\begin{matrix}
U_k\\
V_k
\end{matrix}\right),
\end{equation}
where $(U_k,V_k)^\text{T}$ are the quasiparticle wave functions, $E_k$ is the quasiparticle energy, and $\lambda_\tau$ is the Fermi surface ($\tau = \mathrm{n}/\mathrm{p}$ for neutrons or protons).
The Dirac Hamiltonian $h_D$ in coordinate space reads 
\begin{equation}
h_D(\bm{r})=\bm{\alpha}\cdot\bm{p}+V(\bm{r})+\beta[M+S(\bm{r})],
\end{equation}
with the scalar $S(\bm r)$ and vector $V(\bm r)$ potentials
\begin{equation}
S(\bm r) = \alpha_S \rho_S + \beta_S \rho^2_S + \gamma_S \rho^3_S + \delta_S \Delta \rho_S,
\end{equation}
\begin{equation}
V(\bm r) = \alpha_V \rho_V + \gamma_V \rho^3_V + \delta_V \Delta \rho_V + eA^0 + \alpha_{TV}\tau_3\rho_3 + \delta_{TV} \Delta \tau_3 \rho_3,
\end{equation}
constructed by the local densities:
\begin{equation}
\rho_S(\bm r) = \sum_{k>0} V^\dag_k (\bm r) \gamma_0 V_k (\bm r),
\end{equation}
\begin{equation}
\rho_V(\bm r) = \sum_{k>0} V^\dag_k (\bm r) V_k (\bm r),
\end{equation}
\begin{equation}
\rho_3(\bm r) = \sum_{k>0} V^\dag_k (\bm r) \tau_3 V_k (\bm r).
\end{equation}
Here the no-sea approximation is adopted, i.e., the summation runs over the states in the Fermi sea only.

The pairing potential $\Delta$ reads 
\begin{equation}\label{Delta}
\Delta(\bm r_1,\bm r_2) = V^{\mathrm{pp}}(\bm r_1,\bm r_2)\kappa(\bm r_1,\bm r_2),
\end{equation}
with the density dependent zero-range pairing force $V^{\mathrm{pp}}$,
\begin{equation}\label{pair}
V^{\mathrm{pp}}(\bm r_1,\bm r_2)= V_0 \frac{1}{2}(1-P^\sigma)\delta(\bm r_1-\bm r_2)\left(1-\frac{\rho(\bm r_1)}{\rho_{\mathrm{sat}}}\right),
\end{equation}
and the pairing tensor $\kappa = V^* U^\text{T}$.

The deformed RHB equations in the DRHBc theory are solved in a DWS basis which includes states both in the Fermi sea and in the Dirac sea~\cite{Zhou2003PRC}. The radial wave functions of the DWS basis have a proper asymptotic behavior at large radius $r$. The DRHBc theory is therefore capable of including the contributions of the continuum and describing exotic nuclei with large spatial extensions.

For axially deformed nuclei, the potentials and densities are expanded by the Legendre polynomials, 
\begin{equation}\label{legendre}
f(\bm r)=\sum_{\lambda} f_{\lambda}(r)P_{\lambda}(\cos\theta),~~\lambda=0,2,4,\cdots
\end{equation}
where $\lambda$ is the order of a Legendre polynomial $P_\lambda$, and $f_\lambda$ is the corresponding radial component calculated from  
\begin{equation}\label{leg_f}
f_{\lambda}(r)=\frac{2\lambda+1}{4\pi}\int d\Omega f(\bm{r}) P_\lambda (\Omega).
\end{equation}
As the spatial reflection symmetry is assumed, the Legendre polynomials are limited by even order. If the triaxial deformation degree of freedom is included, the potentials and densities should be expanded by spherical harmonics~\cite{Zhang2023PRCL}.

For an odd-$A$ or odd-odd nucleus, the blocking effect of the unpaired nucleon(s) needs to be considered~\cite{NMBP1980}. The blocked orbital will break the time reversal symmetry and yield the nucleon currents, which would lead to a surge in the computational cost, especially for large-scale calculations. Therefore, the equal filling approximation (EFA)~\cite{Li2012CPL,Martin2008PRC} is adopted, which conserves the time-reversal symmetry and avoids time-consuming computation. In Ref.~\cite{Schunck2010PRC}, it is demonstrated that EFA is an excellent approximation to the strict solution for odd nuclei.

As the translational and rotational invariances are violated in the DRHBc theory, the center-of-mass (c.m.) and rotational (rot) correction energies for the ground-state are microscopically calculated by
\begin{equation}
E_\text{c.m.}=-\frac{1}{2mA}\langle \hat{\bm{P}}^2 \rangle
\end{equation}
and
\begin{equation}
E_\text{rot}=-\frac{1}{2\mathscr{I}}\langle \hat{\bm{J}}^2 \rangle,
\end{equation}
with $A$ the mass number, $\hat{\bm{P}}$ the total momentum in the c.m. frame, $\mathscr{I}$ the moment of inertia calculated by the Inglis-Belyaev formula~\cite{Inglis1956PR,Beliaev1961PR}, and $\hat{\bm{J}}$ the total angular momentum.
Since the EFA is adopted here, the ground-state wave function for an odd-$A$ nucleus in the canonical basis has a similar form as an even-even nucleus.
As the center-of-mass and rotational correction energies are calculated in the canonical basis, the equations for calculating $E_\text{c.m.}$ and $E_\mathrm{rot}$ for an odd nucleus are similar to those for an even-even one.


\section{Numerical details}\label{numerical}

The numerical details for the DRHBc mass table calculation of even-even nuclei have been suggested in Ref.~\cite{Zhang2020PRC} after systematic convergence checks. 
In Ref.~\cite{Pan2022PRC}, it is demonstrated that those numerical details are valid for odd-$A$ and odd-odd nuclei as well. To determine the ground state of an odd-$A$ or odd-odd nucleus, an automatic blocking procedure~\cite{Pan2022PRC} is implemented for reducing the computational cost.

For completeness, in the following we summarize the numerical details in the present DRHBc mass table calculations for even-$Z$ nuclei. 

\begin{itemize}
  \item [$\bullet$] The density functional PC-PK1~\cite{Zhao2010PRC}, which provides one of the best density-functional descriptions for nuclear masses~\cite{Zhao2012PRCmass,Zhang2014FP,Lu2015PRC,Zhang2021PRC}, is employed.
  \item [$\bullet$] In the pairing channel, the density-dependent zero-range pairing force in Eq.~(\ref{pair}) is adopted, where the pairing strength $V_0=-325~\mathrm{MeV~fm}^3$ and the saturation density $\rho_{\mathrm{sat}}=0.152~\mathrm{fm}^{-3}$ together with a pairing window of $100$ MeV, which reproduce well the odd-even mass differences for calcium and lead isotopes~\cite{Zhang2020PRC}.
  \item [$\bullet$] The energy cutoff for the DWS basis in the Fermi sea $E^+_{\mathrm{cut}}=300$ MeV, which guarantees the convergence accuracy of 0.01~MeV for the total energies of doubly magic nuclei $^{40}$Ca, $^{100}$Sn, and $^{208}$Pb~\cite{Zhang2020PRC}, and the deformed heavy odd-$A$ nucleus $^{301}$Th~\cite{Pan2022PRC}.
  \item [$\bullet$] The angular momentum cutoff for the DWS basis $J_{\max}=23/2~\hbar$, which guarantees the convergence accuracy of $0.01\%$ for the total energies of the deformed heavy nuclei $^{300}$Th~\cite{Zhang2020PRC} and $^{301}$Th~\cite{Pan2022PRC}.
  \item [$\bullet$] The number of the DWS basis states in the Dirac sea is the same as that in the Fermi sea~\cite{Zhou2003PRC,Zhou2010PRC,Li2012PRC}.
  \item [$\bullet$] The Legendre expansion truncations in Eq.~(\ref{legendre}) are chosen as $\lambda_{\max}=6$, $8$, and $10$ for nuclei with $8\le Z\le 70$, $72\le Z\le 100$, and $102\le Z\le 120$, respectively~\cite{Zhang2020PRC,Zhang2021PRC}. This guarantees the convergence accuracy of $0.03\%$ for the total energies of $^{20}$Ne and $^{112}$Mo, and of $0.01\%$ for the total energies of $^{300}$Th and $^{301}$Th, when their deformations are constrained to be $\beta_2=0.6$~\cite{Zhang2020PRC,Pan2022PRC}.
  \item [$\bullet$] The automatic blocking procedure~\cite{Pan2022PRC} is implemented for reducing the computational cost, in which the orbital with the lowest quasiparticle energy is blocked during the iteration. If the automatic blocking procedure doesn't work, i.e., the iteration does not converge to a specific blocked orbital, the orbital-fixed blocking procedure is used, where the orbitals near the Fermi surface are blocked separately and the result with the lowest energy is identified as the ground state. 
 
\end{itemize}

\section{Results and Discussion}\label{results}

\subsection{Nuclear masses}
\begin{figure}[htbp]
  \centering
  \includegraphics[width=1.0\textwidth]{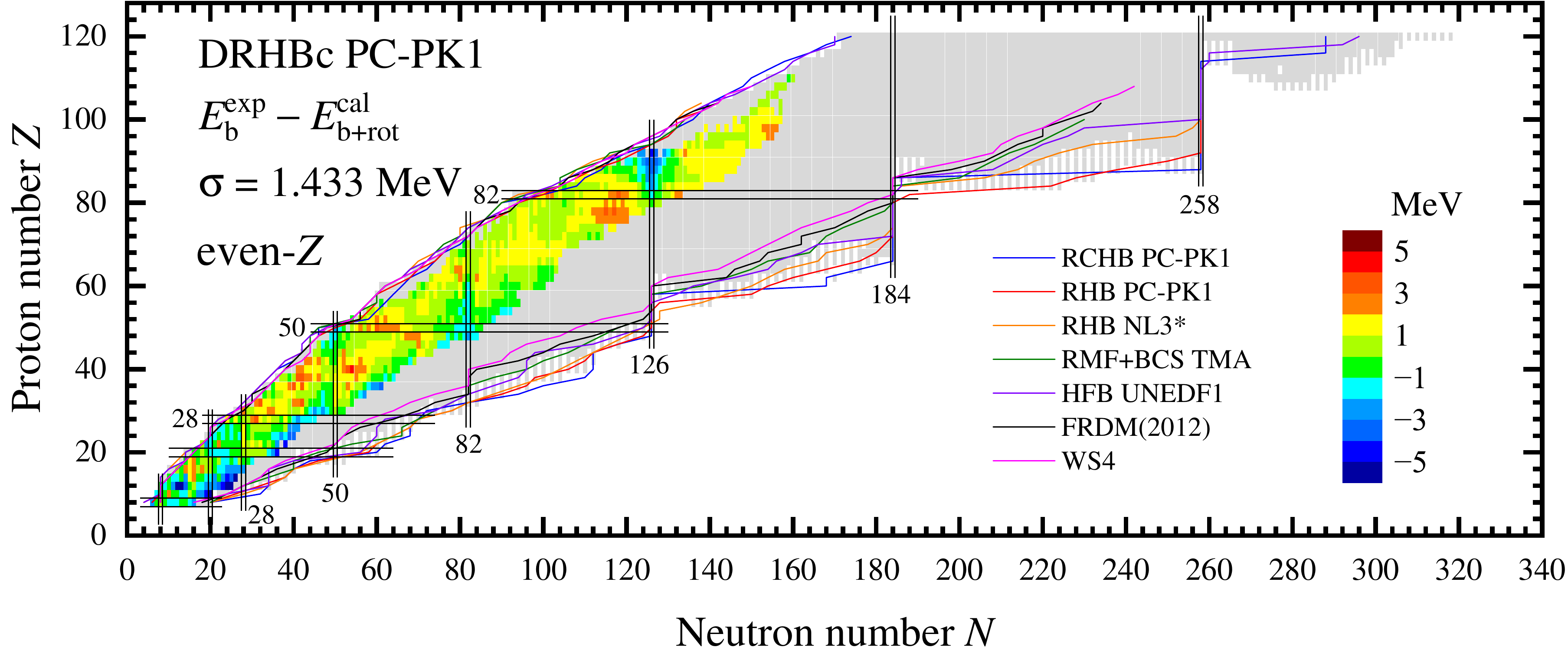}
  \caption{(Color online) 4829 bound even-$Z$ nuclei from O ($Z = 8$) to $Z = 120$ predicted by the DRHBc theory with PC-PK1. For the 1244 even-$Z$ nuclei with mass measured, the binding energy differences between the data~\cite{AME2020(3)} and the DRHBc calculations (with rotational correction energy included) are scaled by colors. The nucleon drip lines predicted by different mass tables, including RCHB with PC-PK1~\cite{Xia2018ADNDT}, RHB with PC-PK1~\cite{Yang2021PRC} and with NL3*~\cite{Agbemava2014PRC}, RMF+BCS with TMA~\cite{Geng2005PTP}, HFB with UNEDF1~\cite{Erler2012Nature}, FRDM(2012)~\cite{Moller2016ADNDT}, and WS4~\cite{Wang2014PLB}, are plotted for comparison.}
\label{fig1}
\end{figure}

Systematic calculations have been performed for all even-$Z$ odd-$N$ nuclei from $Z = 8$ to $Z = 120$ from the proton drip line to the neutron drip line. Together with the results for even-even nuclei compiled in Ref.~\cite{Zhang2022mass}, the ground-state properties of all even-$Z$ nuclei are summarized in Table~\ref{tab1}. The mass number $A$, neutron number $N$, binding energy $E_{\mathrm{b}}^{\mathrm{cal}}$, binding energy including rotational correction energy $E^{\mathrm{cal}}_{\mathrm{b}+\mathrm{rot}}$, two-neutron separation energy $S_\mathrm{2n}$, 
two-proton separation energy $S_\mathrm{2p}$, one-neutron separation energy $S_\mathrm{n}$, neutron rms radius $R_\mathrm{n}$, proton rms radius $R_\mathrm{p}$, matter rms radius $R_m$, charge radius $R_{\mathrm{ch}}$, neutron quadrupole deformation $\beta_\mathrm{2n}$, proton quadrupole deformation $\beta_\mathrm{2p}$, total quadrupole deformation $\beta_{2}$, neutron Fermi surface $\lambda_\mathrm{n}$, and proton Fermi surface $\lambda_\mathrm{p}$ are listed. The available data for binding energies~\cite{AME2020(1),AME2020(2),AME2020(3)} and charge radii~\cite{Angeli2013ADNDT,Li2021ADNDT} are also listed for comparison. In total 4829 even-$Z$ nuclei from O ($Z = 8$) to $Z = 120$ are predicted to be bound, in which 2584 (53.5 \%) are even-even nuclei and 2245 (46.5 \%) are even-$Z$ odd-$N$ ones. In each isotopic chain, all nuclei from the proton drip line to the two-neutron drip line are listed in Table~\ref{tab1} for data completeness. For guidance, two more neutron-deficient and two more neutron-rich unbound nuclei beyond drip lines are also included. If the calculated separation energies ($S_\mathrm{2n}$, $S_\mathrm{2p}$, $S_\mathrm{n}$) of unbound nuclei are negative and/or the Fermi surfaces are positive, these quantities are underlined for clarity. Since only even-$Z$ nuclei are considered in this paper, in the following discussion we refer to even-$Z$ even-$N$ and even-$Z$ odd-$N$ nuclei as even-even and even-odd nuclei, respectively.

Figure~\ref{fig1} illustrates the nuclear landscape of even-$Z$ nuclei from O ($Z = 8$) to $Z = 120$ explored by the DRHBc theory with PC-PK1. On the neutron-rich side, some even-odd nuclei are unbound due to the one-neutron emission. 
For the peninsula of stability with 34 even-even nuclei at $Z=108,110,112$ predicted in Ref.~\cite{Zhang2021PRC}, 17 additional even-odd nuclei are predicted.
Among the total 4829 bound nuclei, the mass data of 1244 nuclei (634 even-even and 610 even-odd nuclei) are available~\cite{AME2020(3)}. For these nuclei, the binding energy differences between the experimental values $E^{\mathrm{exp}}_{\mathrm{b}}$ and the calculated ones $E^{\mathrm{cal}}_{\mathrm{b}+\mathrm{rot}}$ are scaled by colors in the figure. The overall agreement between experimental and calculated binding energies remains satisfactory after including even-odd nuclei; for the 1244 even-$Z$ nuclei the rms deviation $\sigma$=1.433 MeV, which is slightly better than the $\sigma$=1.518 MeV for the even-even nuclei~\cite{Zhang2022mass}.

\subsection{Nucleon separation energies}

Since the DRHBc calculations have been extended to all even-$Z$ nuclei, the two-neutron (two-proton) separation energy $S_\mathrm{2n}$ ($S_\mathrm{2p}$) and the one-neutron separation energy $S_\mathrm{n}$ can be calculated,
\begin{gather}
S_\mathrm{2n}(Z,N) = E_{\mathrm{b}}(Z,N) - E_{\mathrm{b}}(Z,N-2),\\
S_\mathrm{2p}(Z,N) = E_{\mathrm{b}}(Z,N) - E_{\mathrm{b}}(Z-2,N),\\
S_\mathrm{n}(Z,N) = E_{\mathrm{b}}(Z,N) - E_{\mathrm{b}}(Z,N-1).
\end{gather}
These quantities provide information on whether a nucleus is stable against one-neutron or two-nucleon emission, and thus define the one-neutron and two-nucleon drip lines.
In this paper, a nucleus is considered bound only if its one-neutron, two-nucleon, and multi-nucleon separation energies are all positive and its neutron and proton Fermi surfaces are negative.
For each isotopic chain, the two-neutron drip line and the two-proton drip line are defined as $S_\mathrm{2n}=0$ and $S_\mathrm{2p}=0$, respectively. Similarly, the location where $S_\mathrm{n}=0$ defines the one-neutron drip line.

\subsubsection{Two-neutron separation energies}\label{tns}
\begin{figure}[htbp]
  \centering
  \includegraphics[width=1.0\textwidth]{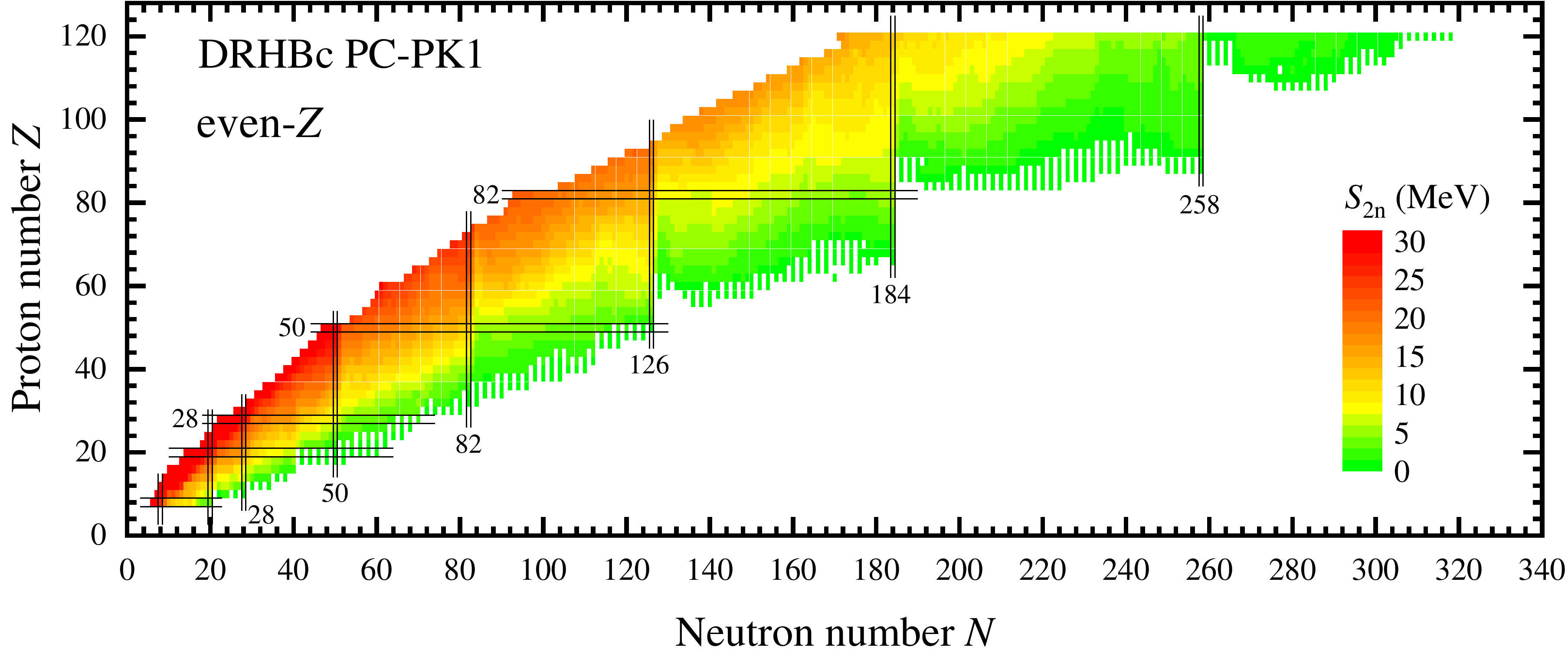}
  \caption{(Color online) Two-neutron separation energies of bound even-$Z$ nuclei with $8 \le Z \le 120$ in the DRHBc calculations with PC-PK1 scaled by colors.}
\label{fig2}
\end{figure}

In Fig.~\ref{fig2}, two-neutron separation energies $S_\mathrm{2n}$ of the bound even-$Z$ nuclei predicted by the DRHBc theory with PC-PK1 are shown.
From a global view, the transition of $S_\mathrm{2n}$ between the even-even and even-odd isotopes is smooth in most cases, and $S_\mathrm{2n}$ is large near the proton drip line and close to zero near the neutron drip line.
For a given isotonic chain, $S_\mathrm{2n}$ increases almost monotonically with the increase of proton number. For a given isotopic chain, $S_\mathrm{2n}$ decreases almost monotonically with the increase of neutron number.
The drastic declines at the known neutron magic numbers $20, 28, 50, 82$, and $126$, and the predicted ones $184$ and $258$ in the superheavy mass region~\cite{Zhang2005NPA,Li2014PLB} are reproduced well by the DRHBc theory.
It is noted that $S_\mathrm{2n}$ of the even-odd nuclei with one neutron more than magic numbers are close to the average of their neighboring even-even nuclei.

In Ref.~\cite{Zhang2022mass}, the number of even-even nuclei in different ranges of $S_\mathrm{2n}$ has been discussed. There are 41 even-even nuclei with $S_\mathrm{2n} \ge 30$ MeV, 103 ones with 21 MeV $\le S_\mathrm{2n} < 30$ MeV, 501 ones with 12 MeV $\le S_\mathrm{2n} < 21$ MeV, 1168 ones with 3 MeV $\le S_\mathrm{2n} < 12$ MeV, and 770 ones with $S_\mathrm{2n} < 3$ MeV.
Correspondingly, the numbers of even-odd nuclei in these ranges are respectively 35, 113, 491, 1166, and 440. 
It can be seen that in the first four ranges, the numbers of even-odd nuclei are close to those of even-even ones.
However, in the range of $S_\mathrm{2n} < 3$ MeV, the number of even-odd nuclei is much smaller than that of even-even ones.
This is because many weakly bound even-even nuclei are beyond the one-neutron drip line.
In addition, it should be noted that there are 278 weakly bound even-$Z$ nuclei with $S_\mathrm{2n} \le 1$ MeV. Among them, there are 253 even-even nuclei that are extremely neutron-rich and lie even beyond the neutron drip lines predicted by other models. There are only 25 even-odd nuclei with $S_\mathrm{2n} \le 1$ MeV and 16 of them are located in the superheavy mass region with $110 \le Z \le 120$.
For these weakly bound even-$Z$ nuclei, because the neutron Fermi surface approaches the continuum threshold, pairing correlations could scatter nucleons from bound states to resonant ones in the continuum and, thus, provide a significant coupling between the continuum and bound states, which might affect the location of the drip line~\cite{Xia2018ADNDT}.
In addition, the nearly vanishing $S_\mathrm{2n}$ around the neutron drip line might indicate halo phenomena~\cite{Meng2006PPNP}. 
Detailed analysis of the neutron rms radii and single-neutron orbitals around the Fermi surface is interesting for further exploration.

\subsubsection{Two-proton separation energies}
\begin{figure}[htbp]
  \centering 
  \includegraphics[width=1.0\textwidth]{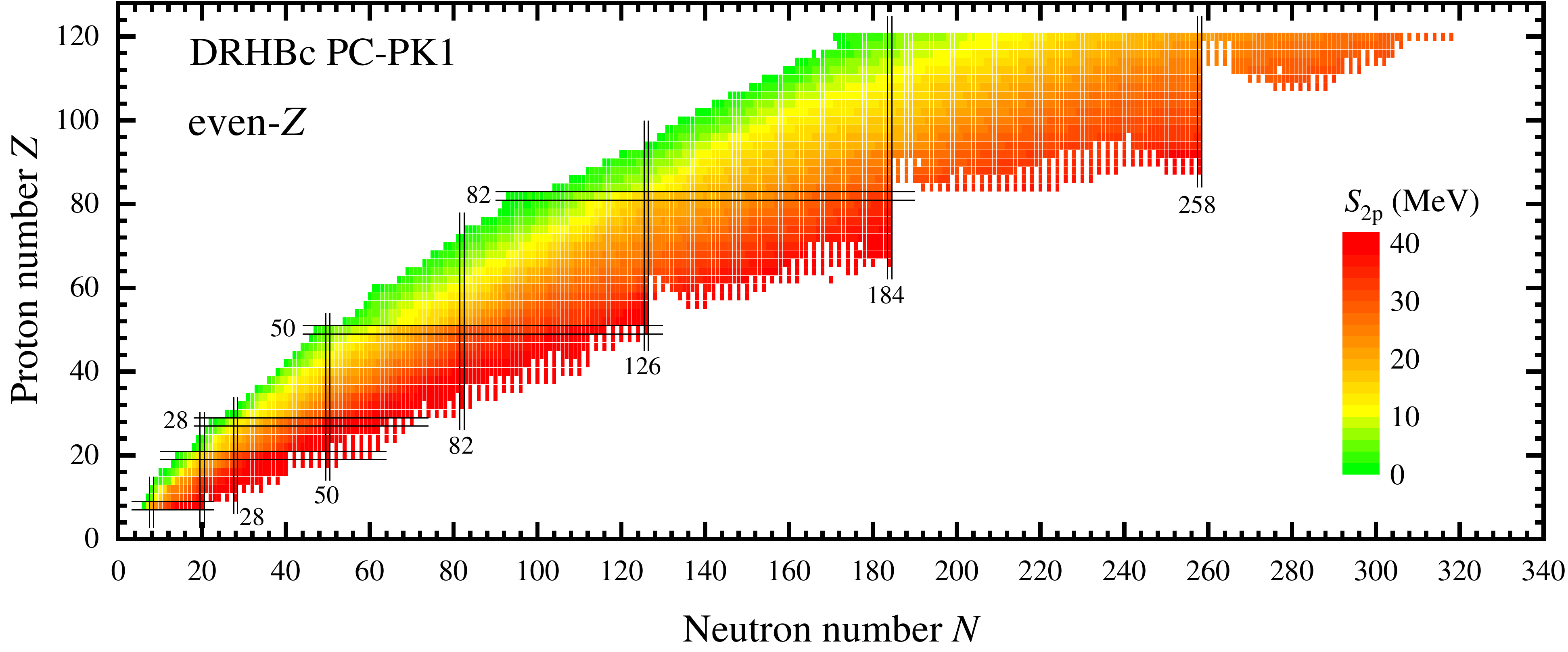}
  \caption{(Color online) Two-proton separation energies of bound even-$Z$ nuclei with $8 \le Z \le 120$ in the DRHBc calculations with PC-PK1 scaled by colors.}
\label{fig3}
\end{figure}

In Fig.~\ref{fig3}, two-proton separation energies $S_\mathrm{2p}$ of the bound even-$Z$ nuclei predicted by the DRHBc theory with PC-PK1 are scaled by colors.
$S_\mathrm{2p}$ increases almost monotonically with the increase of neutron number for a given isotopic chain. $S_\mathrm{2p}$ decreases almost monotonically with the increase of proton number for a given isotonic chain. 
The drastic declines at proton magic numbers $20, 28, 50$, and $82$ demonstrate that the DRHBc theory reproduces the traditional proton shell closures.

According to the predictions in Ref.~\cite{Zhang2022mass}, there are 232 even-even nuclei with $S_\mathrm{2p} \ge 40$ MeV, 611 ones with 30 MeV $\le S_\mathrm{2p} < 40$ MeV, 754 ones with 20 MeV $\le S_\mathrm{2p} < 30$ MeV, 566 ones with 10 MeV $\le S_\mathrm{2p} < 20$ MeV, and 421 ones with $S_\mathrm{2p} < 10$ MeV.
The corresponding numbers of even-odd nuclei in these ranges are respectively 86, 455, 719, 569, and 416. Except for the neutron-rich region with $S_\mathrm{2p} \ge $ 30 MeV, the numbers of even-odd nuclei in all the other $S_\mathrm{2p}$ intervals are very close to those of even-even ones, which is consistent with the result in Section~\ref{tns} that the number of even-odd nuclei with $S_\mathrm{2n} < 3$ MeV is much smaller than that of even-even ones.
There are 14 even-odd and 17 even-even weakly bound nuclei with $S_\mathrm{2p} \le 1$ MeV.
Similar to the conclusion in Ref.~\cite{Zhang2022mass}, the number of even-$Z$ nuclei with $S_\mathrm{2p} \le 1$ MeV is much smaller than that with $S_\mathrm{2n} \le 1$ MeV. 
This indicates that the continuum effects are suppressed on the proton-rich side due to the existence of the Coulomb barrier.

\subsubsection{One-neutron separation energies} 
\begin{figure}[htbp]
  \centering
  \includegraphics[width=1.0\textwidth]{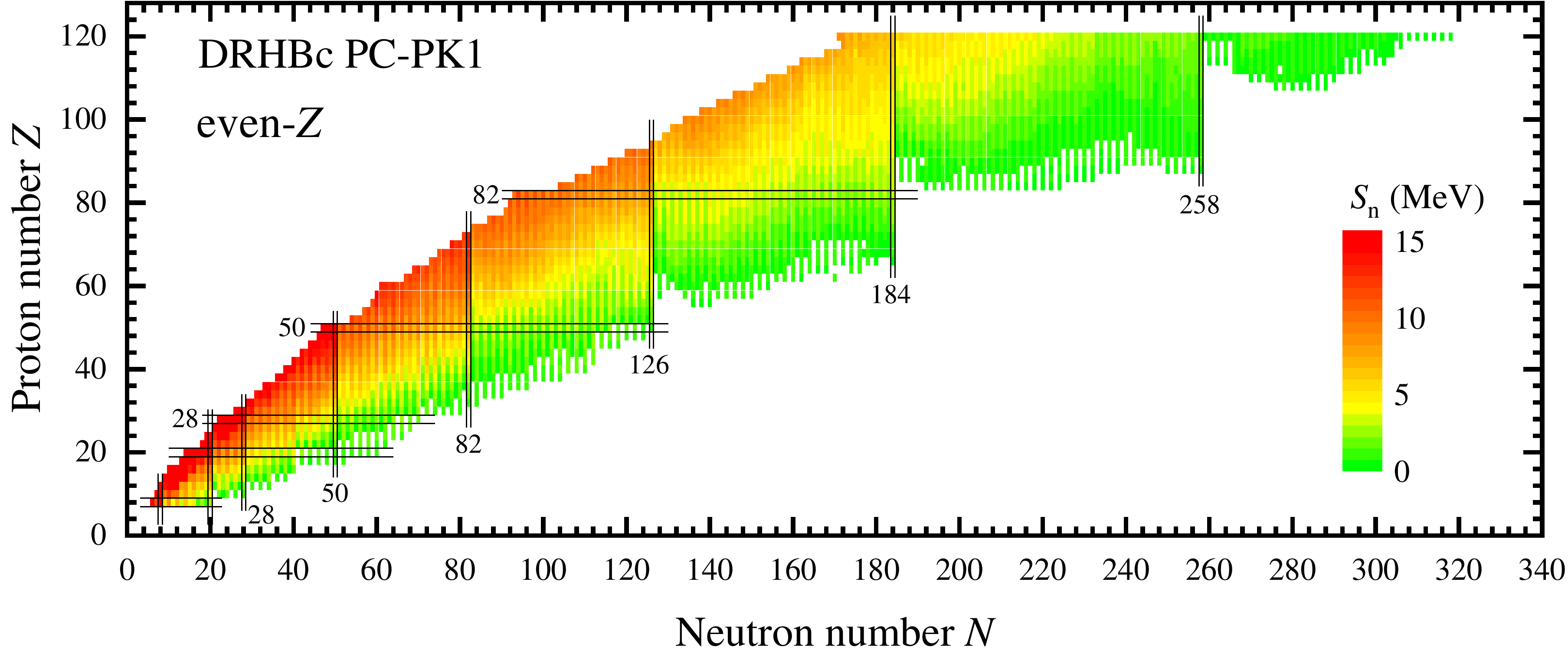}
  \caption{(Color online) One-neutron separation energies of bound even-$Z$ nuclei with $8 \le Z \le 120$ in the DRHBc calculations with PC-PK1 scaled by colors.}
\label{fig4}
\end{figure}

In Fig.~\ref{fig4}, one-neutron separation energies $S_\mathrm{n}$ of the bound even-$Z$ nuclei predicted by the DRHBc theory with PC-PK1 are shown.
Similar to $S_\mathrm{2n}$, $S_\mathrm{n}$ is large near the proton drip line and close to zero near the neutron drip line.
For a given isotonic chain, $S_\mathrm{n}$ increases with the increase of proton number. For a given isotopic chain, $S_\mathrm{n}$ generally decreases with the increase of neutron number, while significant odd-even staggering exists, i.e., the $S_\mathrm{n}$ of an even-odd nucleus is obviously lower than those of its neighboring even-even ones. Therefore, on the neutron-rich side, for an even-even nucleus with a positive $S_\mathrm{n}$, its neighboring even-odd ones may have a negative $S_\mathrm{n}$, which leads to the fact that the one-neutron and two-neutron drip line locations are different in most isotopic chains. 
Besides, it is noted that at neutron magic numbers $N = 20, 28, 50, 82, 126, 184$, and $258$, $S_\mathrm{n}$ shows abrupt changes.

Comparing Figs.~\ref{fig2} and~\ref{fig4}, one can find that $S_\mathrm{n}$ is close to half of $S_\mathrm{2n}$ for most nuclei.
There are 69 nuclei (48 even-even and 21 even-odd ones) with $S_\mathrm{n} \ge 15$ MeV located near the proton drip line, 230 proton-rich nuclei (154 even-even and 76 even-odd ones) with 10.5 MeV $\le S_\mathrm{n} < 15$ MeV, 961 nuclei (512 even-even and 449 even-odd ones) with 6 MeV $\le S_\mathrm{n} < 10.5$ MeV near the $\beta$-stability line, 2374 nuclei (1326 even-even and 1048 even-odd ones) with 1.5 MeV $\le S_\mathrm{n} < $ 6 MeV, and 1194 nuclei (543 even-even and 651 even-odd ones) with $S_\mathrm{n} < $ 1.5 MeV near or even beyond the one-neutron drip line.
It is found that there are 289 nuclei with $S_\mathrm{n} < $ 0.5 MeV which are weakly bound against one-neutron emission.
270 of them are even-odd nuclei close to the one-neutron drip line, while only 19 of them are even-even ones. 
This is because $S_\mathrm{n}$ of most of even-even nuclei is larger than 0.5 MeV even close to the one-neutron or two-neutron drip line, which reflects that in the extremely neutron-rich region even-even nuclei generally have larger binding energies than their adjacent even-odd ones.
It is also noted that most of even-even nuclei with very small $S_\mathrm{n}$ are located in the region with $110 \le Z \le 120$.
Besides, these even-odd nuclei with very small $S_\mathrm{n}$ are also candidates for one-neutron halo nuclei and further analysis on neutron rms radii and single-particle levels can be performed.

\subsection{Comparison with other predictions}\label{secC}
\subsubsection{Masses and separation energies}
\begin{table}
\centering
\caption{The rms deviations for the binding energies, two-neutron separation energies, one-neutron separation energies, and two-proton separation energies in the DRHBc calculations with PC-PK1 with respect to the AME2020 data~\cite{AME2020(3)} in the unit of MeV. The results of other relativistic and non-relativistic density functional calculations are also listed for comparison. All bound even-$Z$ nuclei with available experimental data are included in the calculation.}
\begin{tabular}{cccccccc}
\hline
Model &  Density functional & ~~$\sigma(E_{\mathrm{b}})$~~ & ~~$\sigma(S_\mathrm{2n})$~~ & ~~$\sigma(S_\mathrm{n})$~~ & ~~$\sigma(S_\mathrm{2p})$~~ & Data number & Reference \\
\hline
DRHBc$^{\mathrm{w/o}~E_{\mathrm{rot}}}$ &  PC-PK1 & \textbf{2.562} & \textbf{0.953} & \textbf{0.749} & \textbf{0.929} & 1244 & This work \\
DRHBc$^{\mathrm{w/} ~~E_{\mathrm{rot}}}$ &  PC-PK1 & \textbf{1.433} & \textbf{0.989} & \textbf{0.774} & \textbf{1.046} & 1244 & This work \\
RCHB &  PC-PK1 & 8.082 & 1.492 & 0.835 & 1.561 & 1232 & ~~\cite{Xia2018ADNDT} \\
RMF+BCS & TMA & 2.063 & 0.897 & 0.730 & 1.114 & 1249 &~~\cite{Geng2005PTP}\\
\hline
HFB &  SkM$^{*}$ & 7.248 & 1.134 & 0.693 & 1.781 & 1239 &~~\cite{Erler2012Nature} \\
HFB &  SLy4 & 5.275 & 0.914 & 0.598 & 0.855 & 1249 &~~\cite{Erler2012Nature} \\
HFB &  SV-min & 3.391 & 0.731 & 0.481 & 0.778 & 1248 &~~\cite{Erler2012Nature} \\
HFB &  UNEDF1 & 1.934 & 0.692 & 0.520 & 0.780 & 1250 &~~\cite{Erler2012Nature} \\
\hline
\end{tabular}
\label{tab2}
\end{table}

For a quantitative comparison with previous works, the rms deviations of binding energies $\sigma(E_{\mathrm{b}})$, two-neutron separation energies $\sigma(S_\mathrm{2n})$, one-neutron separation energies $\sigma(S_\mathrm{n})$, and two-proton separation energies $\sigma(S_\mathrm{2p})$ for the present calculations with respect to the data available from AME2020~\cite{AME2020(3)} are listed in Table~\ref{tab2}, together with those for several previous density functional calculations.
Because the numbers of bound nuclei may differ by models, the data numbers for extracting the rms deviations are also listed in Table~\ref{tab2}.
The rms deviations for even-$Z$ nuclei with and without rotational correction energies are 1.433 MeV and 2.562 MeV, respectively, slightly smaller than those for even-even nuclei with $\sigma = 1.518$ MeV and $2.744$ MeV~\cite{Zhang2022mass}. 
According to Table~\ref{tab2}, the conclusions extracted in Ref.~\cite{Zhang2022mass} stay the same when including the results of even-odd nuclei.
By comparing the DRHBc results with those from other relativistic and nonrelativistic density functional calculations, the present PC-PK1 calculations including rotational correction energies provide a better description for nuclear masses.
The comparison with RCHB results demonstrates the importance of the deformation degrees of freedom in describing the nuclear masses.
The DRHBc results with the rotational correction energies further improve the description for the nuclear masses.

The accuracies for two-nucleon separation energies of most density functionals in Table~\ref{tab2} are around 1 MeV.
The accuracy for two-nucleon separation energies of the RCHB theory is about 1.5~MeV, larger than the others due to the assumed spherical symmetry.
The accuracy of one-neutron separation energies obtained by different density functionals is about 0.5 to 1.0 MeV.
It is noted that the rotational correction does not improve the description of one-neutron and two-nucleon separation energies in the present calculations.
This is because the cranking approximation is not suitable for spherical and weakly deformed nuclei. 
In Ref.~\cite{ZhangXY2023PRC}, with the DRHBc + 2DCH method adopted~\cite{Sun2022CPC}, the rms deviation of $E_\text{rot}$ for Kr and Sr isotopes is reduced from 3.0 MeV to 1.8 MeV with the rotational correction energies, and further reduced to 1.2 MeV with the dynamical correction. Similar improvement can be found for two-neutron separation energies, i.e., the rms deviation of $S_\text{2n}$ are 1.01 MeV, 1.16 MeV, and 0.77 MeV for the DRHBc, the DRHBc with $E_\text{rot}$, and DRHBc+2DCH theories, respectively.
The DRHBc + 2DCH method is also expected to improve the description for even-odd nuclei in future works.

\subsubsection{Limits of the nuclear landscape}

In Fig.~\ref{fig1}, the nucleon drip lines predicted by the DRHBc calculations together with other mass tables, including RCHB with PC-PK1~\cite{Xia2018ADNDT}, RHB with PC-PK1~\cite{Yang2021PRC} and with NL3*~\cite{Agbemava2014PRC}, RMF+BCS with TMA~\cite{Geng2005PTP}, HFB with UNEDF1~\cite{Erler2012Nature}, FRDM(2012)~\cite{Moller2016ADNDT}, and WS4~\cite{Wang2014PLB}, have been plotted.
Due to the Coulomb interaction among protons, the proton drip line does not lie very far away from the valley of stability, and the experimental explorations of the proton drip line have successfully reached up to neptunium ($Z=93$)~\cite{Zhang2019PRL}.
At the same time, as shown in Fig.~\ref{fig1}, the proton drip lines predicted by different mass models are quite close to each other and all roughly consistent with the experimental observations.
This is because the Coulomb barrier suppresses the continuum effects.

On the neutron-rich side, however, the limits of the nuclear landscape are far away from the stability valley. The neutron-rich boundary is known only up to neon ($Z = 10$) experimentally~\cite{Ahn2019PRL}.
From Fig.~\ref{fig1}, it can be seen that the predicted neutron drip lines depend on the theoretical model and/or the employed density functional, and their differences increase with the proton number.
In most regions, it can be found that the deformation effect does not necessarily extend the neutron drip line~\cite{EJ2021IJMPE}.
As discussed in Ref.~\cite{Zhang2022mass}, compared with the neutron drip line in the spherical RCHB theory, the peninsula with $108 \le Z \le 112$ can be found after the inclusion of the deformation. There are also bound even-odd nuclei inside the peninsula. However, most even-odd nuclei are unbound against the one-neutron emission at the edge.
Besides, there are some smaller peninsulas consisted of even-even nuclei $^{192-196}$Ba, $^{192-208}$Ce, $^{232}$Sm, $^{238,240}$Gd, $^{250}$Dy, $^{276-308}$Po, and $^{338-346}$Po.
The formation of these peninsulas is related to deformation effects~\cite{Zhang2021PRC,He2021CPC,Pan2021PRC}.

Compared with the two-neutron drip line, the one-neutron drip line occurs earlier except at neutron magic numbers $N =126, 184$, and $258$.
The neutron drip lines predicted by the macroscopic-microscopic mass models FRDM~\cite{Moller2016ADNDT} and WS4~\cite{Wang2014PLB} are systematically closer to the valley of stability.
The proper treatment of the continuum in the DRHBc theory and the adopted density functional PC-PK1 largely contribute to the much more extended neutron drip lines, as shown in Fig.~\ref{fig1}.

Compared with the two-neutron drip line in DRHBc, RCHB extrapolates two more neutrons for $Z = 24$, 40, and 48, four more neutrons for $Z = 10$, 34, and 36, and fourteen more neutrons for $Z = 60$. RHB + NL3* extrapolates two more neutrons for $Z = 52$ and 54. RHB + PC-PK1 which includes the triaxial deformation degrees of freedom extrapolates four more neutrons for $Z = 82$. Otherwise, the DRHBc calculations with PC-PK1 predict a more extended neutron drip line.
\subsection{Odd-even mass differences}
\begin{figure}[htbp]
  \centering
  \includegraphics[width=1.0\textwidth]{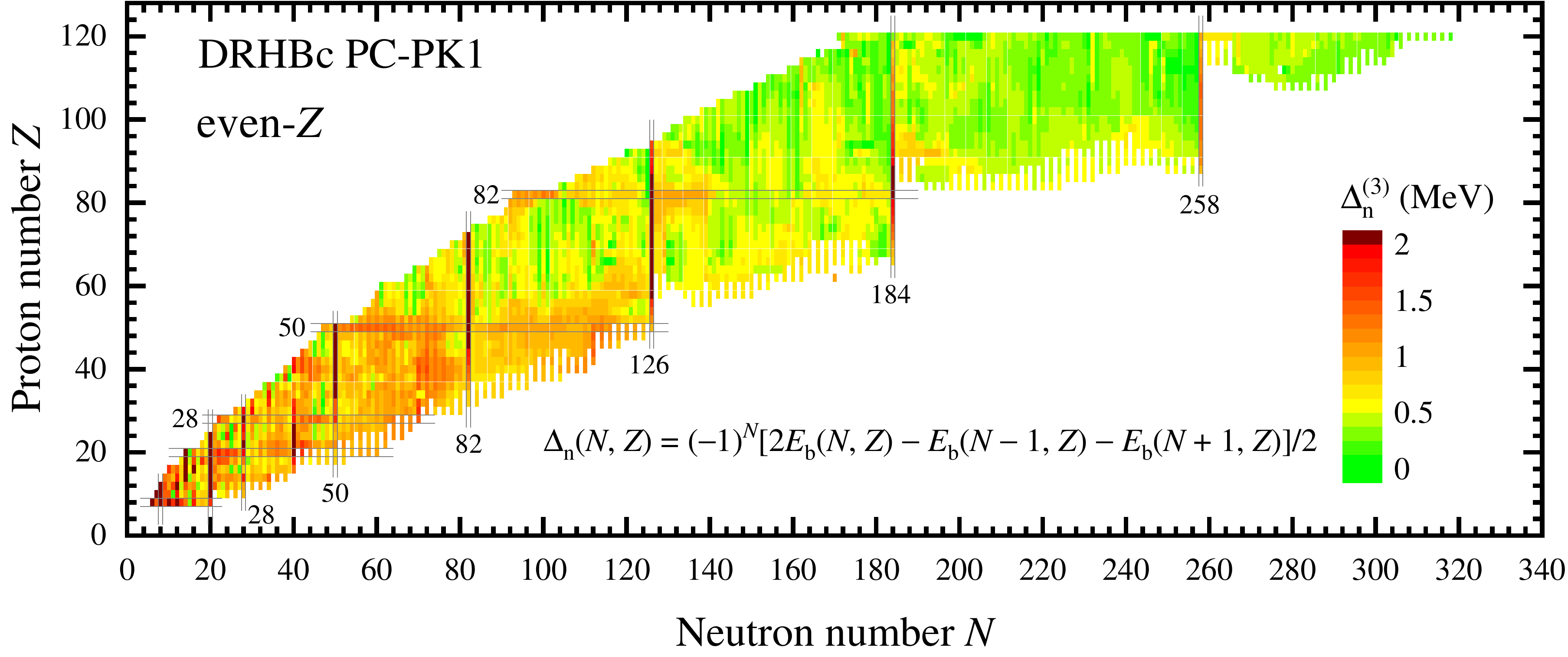}
  \caption{(Color online) Odd-even mass differences of bound even-$Z$ nuclei with $8 \le Z \le 120$ in the DRHBc calculations with PC-PK1 scaled by colors.}
\label{fig5}
\end{figure}

The inclusion of the odd-$A$ nuclei in the DRHBc mass table allows us to extract the odd-even mass difference, which is closely related to the effect of pairing correlations. The odd-even mass difference can be given by the following three-point formula
\begin{gather}
\Delta_\mathrm{n}(N,Z) = \frac{(-1)^N}{2} [2E_{\mathrm{b}}(N,Z)-E_{\mathrm{b}}(N-1,Z)-E_{\mathrm{b}}(N+1,Z)].
\end{gather}
Figure~\ref{fig5} shows the odd-even mass differences $\Delta_\mathrm{n}(N,Z)$ of the bound even-$Z$ nuclei predicted by the DRHBc theory with PC-PK1.

From a global view, there is a general trend that $\Delta_\mathrm{n}$ decreases with the increase of mass number $A$. It can be also noticed that there are apparent peaks at neutron magic numbers $N=20, 28, 50, 82, 126, 184$ and $258$, owing to the shell closure. 
Besides, the nuclei near the limits of the nuclear landscape at the neutron rich side have larger $\Delta_\mathrm{n}$ in general.
This is consistent with the larger neutron pairing energy near the neutron drip line (cf. Fig.~\ref{fig6}), which may be attributed to the high density of single-particle levels near the threshold.
Meanwhile, odd-even staggerings are shown in $\Delta_\mathrm{n}$, i.e., $\Delta_\mathrm{n}$ for even-even nuclei are generally larger than those for their neighboring odd ones. As discussed in Ref.~\cite{Satula1998PRL}, $\Delta_\mathrm{n}$ of even-even nuclei is strongly affected by both nucleonic pairing and the deformed mean field whereas the pairing effect dominates that of odd-$A$ ones. 

From the available experimental binding energies, the $\Delta_\mathrm{n}$ for 1123 even-$Z$ nuclei can be extracted, and the deviation between the experimental and theoretical data is $\sigma = 0.580$ MeV. 
The deviations are $\sigma = 0.624$ MeV for 555 even-even nuclei and $\sigma = 0.532$ MeV for 568 even-odd ones.
When including the rotational correction energies $E_\text{rot}$, however, the rms deviation with respect to the experimental data slightly increases to $\sigma = 0.597$ MeV. One reason for the larger deviation after including $E_\text{rot}$ is that the rotational correction energy is simply calculated using the cranking approximation, which has been discussed in Section~\ref{secC}. Further improvement by the collective Hamiltonian method~\cite{Sun2022CPC,ZhangXY2023PRC} is expected.

\subsection{Pairing energies}
\subsubsection{Neutron pairing energies}
\begin{figure*}[htbp]
  \centering
  \includegraphics[width=1.0\textwidth]{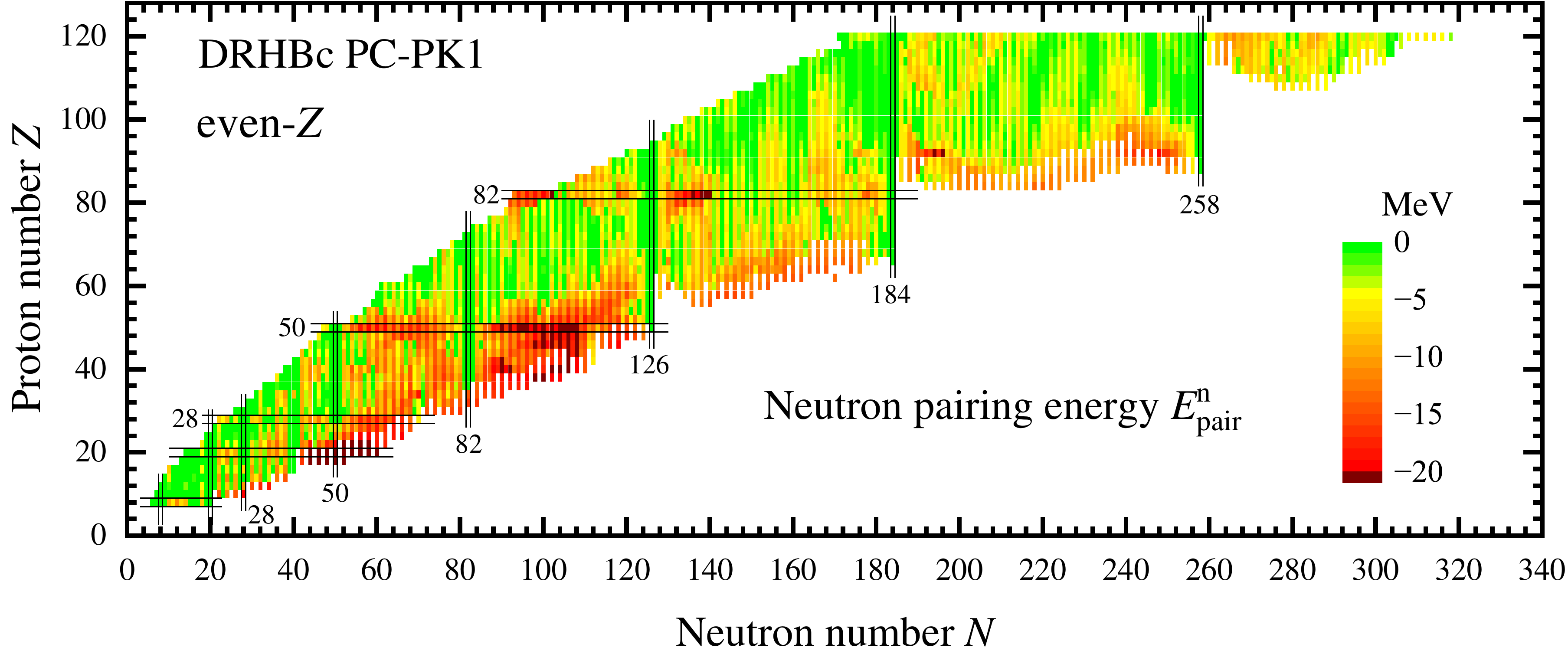}
  \caption{(Color online) Neutron pairing energies of bound even-$Z$ nuclei with $8 \le Z \le 120$ in the DRHBc calculations with PC-PK1 scaled by colors.}
\label{fig6}
\end{figure*}

To examine the pairing correlations from a global view, the pairing energies for even-$Z$ nuclei have been investigated in the DRHBc calculations with PC-PK1.
Figure~\ref{fig6} shows the neutron pairing energies $E_{\mathrm{pair}}^\mathrm{n}$ of bound even-$Z$ nuclei with $8 \le Z \le 120$ scaled by colors.
Near the closed shells $N = 8, 20, 28, 50, 82$ and $126$, it can be clearly seen that the neutron pairing energies become close to zero or even vanish, while generally the maximum values can be found in the middle of the shells.
At $N = 184$ and $258$, the vanished neutron pairing energies agree well with the above discussion for shell closures by $S_\mathrm{2n}$ and $S_\mathrm{n}$.
The odd-even staggerings are shown in $E_{\mathrm{pair}}^\mathrm{n}$, i.e., $E_{\mathrm{pair}}^\mathrm{n}$ for even-odd nuclei is often smaller than their neighboring even-even ones, or even vanishes.
Such odd-even staggering is caused by the unpaired particle, which hinders the scattering of the nucleon pairs.
For light nuclei with $Z<20$ and $N<20$, it can be found that $E_{\mathrm{pair}}^\mathrm{n}$ is very small, especially for even-odd nuclei whose $E_{\mathrm{pair}}^\mathrm{n}$ are very close to 0, partly due to the low density of single-particle levels.
In other regions with very small or even vanished $E_{\mathrm{pair}}^\mathrm{n}$ along a part of an isotonic chain, it may often be related to spherical or deformed subshells, e.g., $N = 40$ in the $Z \sim 20$ region and $N = 162$ in the $Z \sim 110$ region, which will be further discussed in Section~\ref{tng}.
Finally, $E_{\mathrm{pair}}^\mathrm{n}$ for nuclei at $N = 28$ for $Z \sim$ 10, $N = 50$ for $Z \sim $ 20, and $N = 82$ for $Z \sim$ 32 are pronounced, and $E_{\mathrm{pair}}^\mathrm{n}$ for odd nuclei $^{71}$Ti and $^{75}$Cr are respectively $-15$ MeV and $-8$ MeV.
These results indicate the disappearance of the traditional neutron magic numbers 28, 50, and 82 in these neutron-rich nuclei.

\subsubsection{Proton pairing energies}
\begin{figure*}[htbp]
  \centering
  \includegraphics[width=1.0\textwidth]{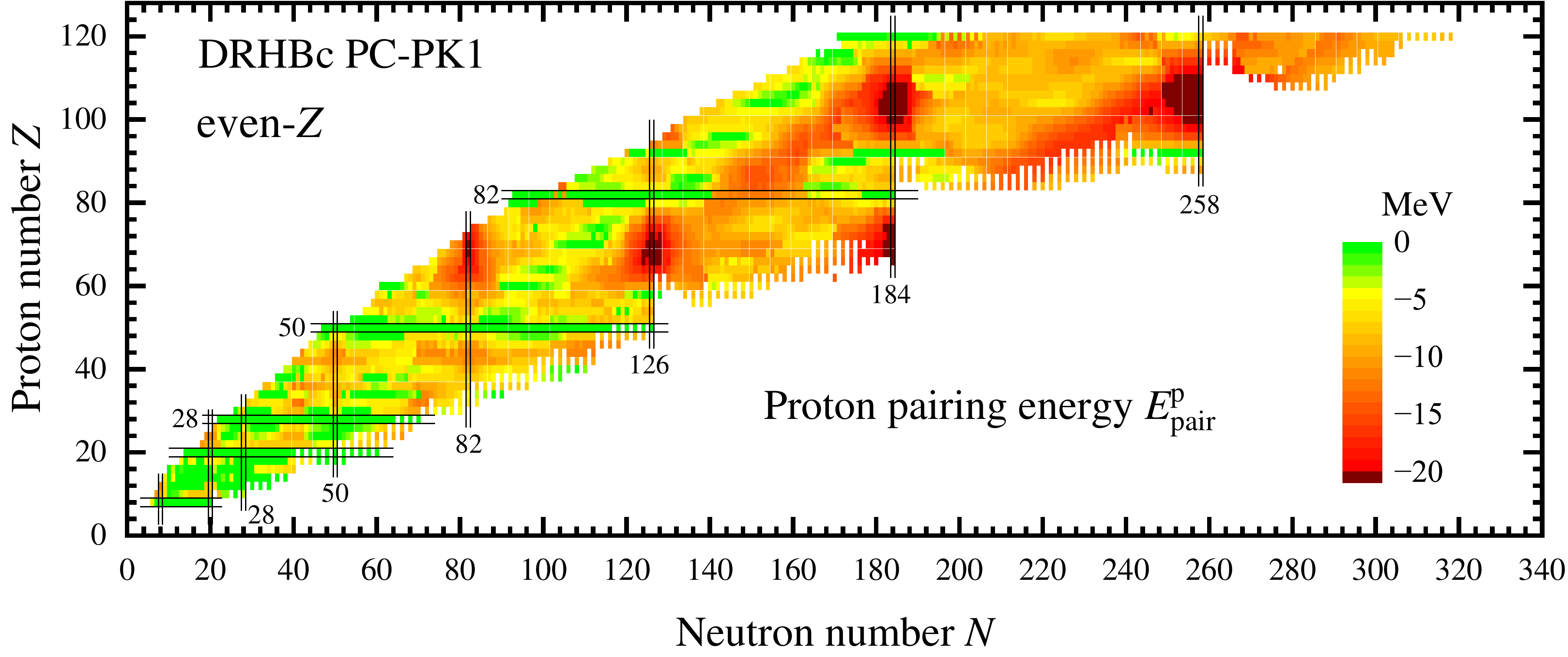}
  \caption{(Color online) Proton pairing energies of bound even-$Z$ nuclei with $8 \le Z \le 120$ in the DRHBc calculations with PC-PK1 scaled by colors.}
\label{fig7}
\end{figure*}

The proton pairing energies $E_{\mathrm{pair}}^\mathrm{p}$ of bound even-$Z$ nuclei with $8 \le Z \le 120$ are presented in Fig.~\ref{fig7}.
No obvious odd-even staggering can be found in $E_{\mathrm{pair}}^\mathrm{p}$, i.e., $E_{\mathrm{pair}}^\mathrm{p}$ for even-odd nuclei is consistent with their neighboring even-even ones.
There are many features similar to the neutron pairing energies, such as that $E_{\mathrm{pair}}^\mathrm{p}$ nearly vanishes for most nuclei near the closed shells $Z = 8, 20, 28, 50,$ and $82$. 
However, it can be seen that $^{185}$Pb and $^{187}$Pb have much larger $E_{\mathrm{pair}}^\mathrm{p}$ compared with their neighboring isotopes.
This is because they are well-deformed with $\beta_2\approx 0.3$, while their neighbors are either spherical or oblate, as seen in Fig.~\ref{fig14}.
This is consistent with the prolate shape confirmed in $^{185}$Pb~\cite{Pakarinen2009PRC}.
For the region near the proton drip line at $Z=120$, $E_{\mathrm{pair}}^\mathrm{p}$ approaches zero or even vanishes for most nuclei.
By increasing the neutron number for $Z=120$, odd-even staggering is shown near $N=200$, which is related to the prolate-oblate shape change, as seen in Fig.~\ref{fig14}. 
Apart from proton magic numbers, there are some nuclei with quite small $E_{\mathrm{pair}}^\mathrm{p}$.
For example, $E_{\mathrm{pair}}^\mathrm{p}$ is very small for light nuclei with $Z<20$ and $N<20$, related to the low density of single-particle levels in this region.
Some small $E_{\mathrm{pair}}^\mathrm{p}$ along a part of an isotopic chain can also be found, e.g., near $N=60$ for $Z=34$ and near $N=184, 258$ for $Z=92$.
As discussed in Ref.~\cite{Zhang2022mass}, the former might be a signal for a deformed proton subshell, and the latter might be related to the spherical nuclei at $Z=92$ which has been considered as a pseudo shell in many relativistic density functionals~\cite{Geng2006CPL}.

\subsection{Two-nucleon gaps}

The two-neutron gap $\delta_\mathrm{2n}$ and the two-proton gap $\delta_\mathrm{2p}$ are respectively defined as
\begin{gather}
\delta_\mathrm{2n}(Z,N) = S_\mathrm{2n}(Z,N) - S_\mathrm{2n}(Z,N+2),\\
\delta_\mathrm{2p}(Z,N) = S_\mathrm{2p}(Z,N) - S_\mathrm{2p}(Z+2,N).
\end{gather}
The peaks in the two-nucleon gaps, which indicate the drastic change of the two-nucleon separation energies, can be regarded as signatures for magic numbers~\cite{Zhang2005NPA,Li2014PLB}.
Therefore, in addition to two-nucleon and one-nucleon separation energies, the two-nucleon gaps are very useful for discovering possible magic numbers and exploring possible subshells.

\subsubsection{Two-neutron gaps}\label{tng}
\begin{figure*}[htbp]
  \centering
  \includegraphics[width=1.0\textwidth]{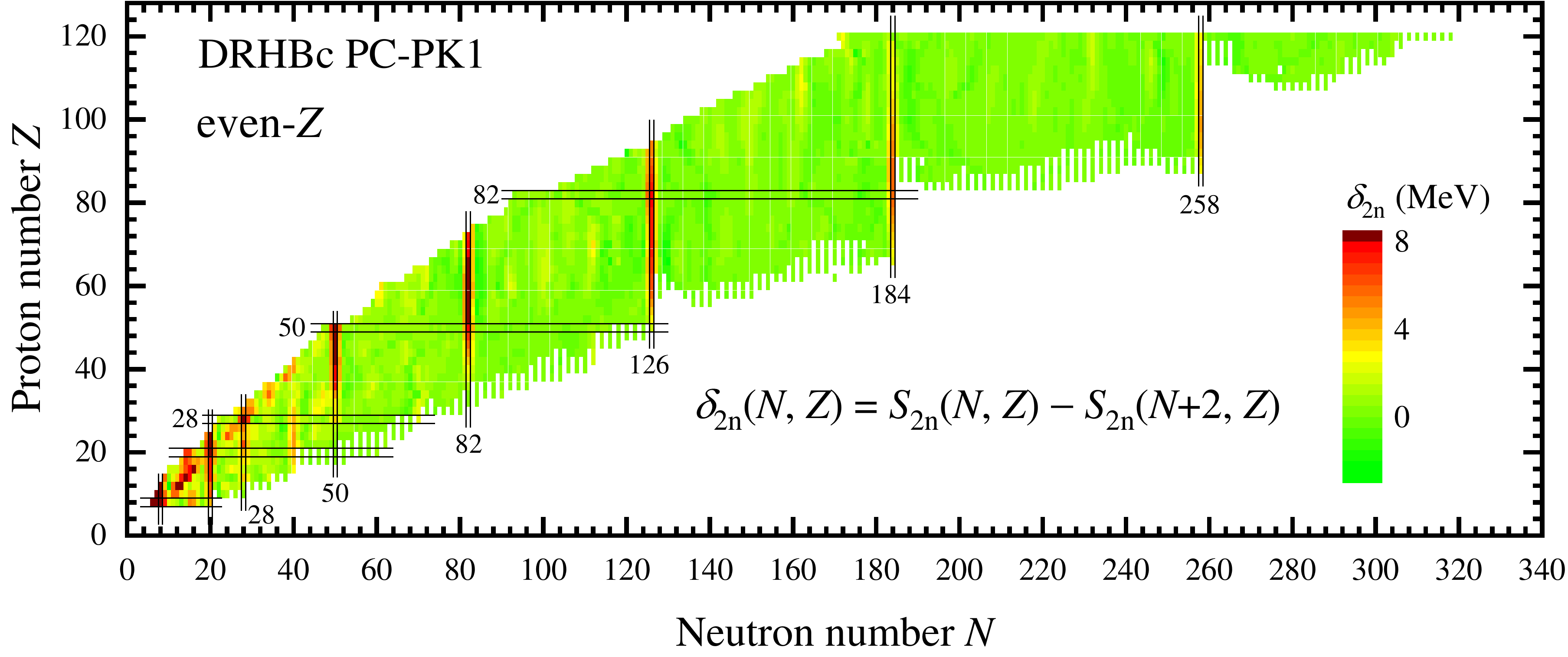}
  \caption{(Color online) Two-neutron gaps $\delta_\mathrm{2n}$ of bound even-$Z$ nuclei with $8 \le Z \le 120$ in the DRHBc calculations with PC-PK1 scaled by colors.}
\label{fig8}
\end{figure*}

Figure~\ref{fig8} shows the two-neutron gaps $\delta_\mathrm{2n}$ of bound even-$Z$ nuclei with $8 \le Z \le 120$ in the DRHBc calculations with PC-PK1.
$\delta_\mathrm{2n}$ for even-odd isotopes present great consistency with those for the even-even ones.
The peaks of $\delta_\mathrm{2n}$ at neutron magic numbers $N = 8, 20, 28, 50, 82, 126, 184$ and $258$ can be clearly seen in Fig.~\ref{fig8}, which is consistent with the conclusion for shell closures discussed above.
The peaks generally become lower with the increase of neutron number, reflecting the weakening of the shell effect in heavy nuclei.
In addition, $\delta_\mathrm{2n}$ of even-odd nuclei adjacent to the neutron magic number is also significant and close to the half of the peak.
This is the reflection of the drastic change of $S_\mathrm{2n}$ for even-even nuclei at neutron magic numbers and their adjacent even-odd nuclei.
As discussed in Ref.~\cite{Zhang2022mass}, the peaks at traditional neutron numbers $N = 28, 50,$ and $82$ become weaker or even disappear near the neutron drip lines, suggesting the quenching or even collapse of the traditional neutron shell closures in the neutron-rich region.
There are some smaller peaks at $N=40$ for $Z \sim 20$ and $N=162$ for $Z \sim 110$, which can be considered as hints for spherical or deformed subshells.
Further confirmation of such subshells needs detailed analysis for deformation, blocked orbital in even-odd nuclei, and the evolution of single-neutron levels with deformation from constrained calculations.

\subsubsection{Two-proton gaps}
\begin{figure*}[htbp]
  \centering  
  \includegraphics[width=1.0\textwidth]{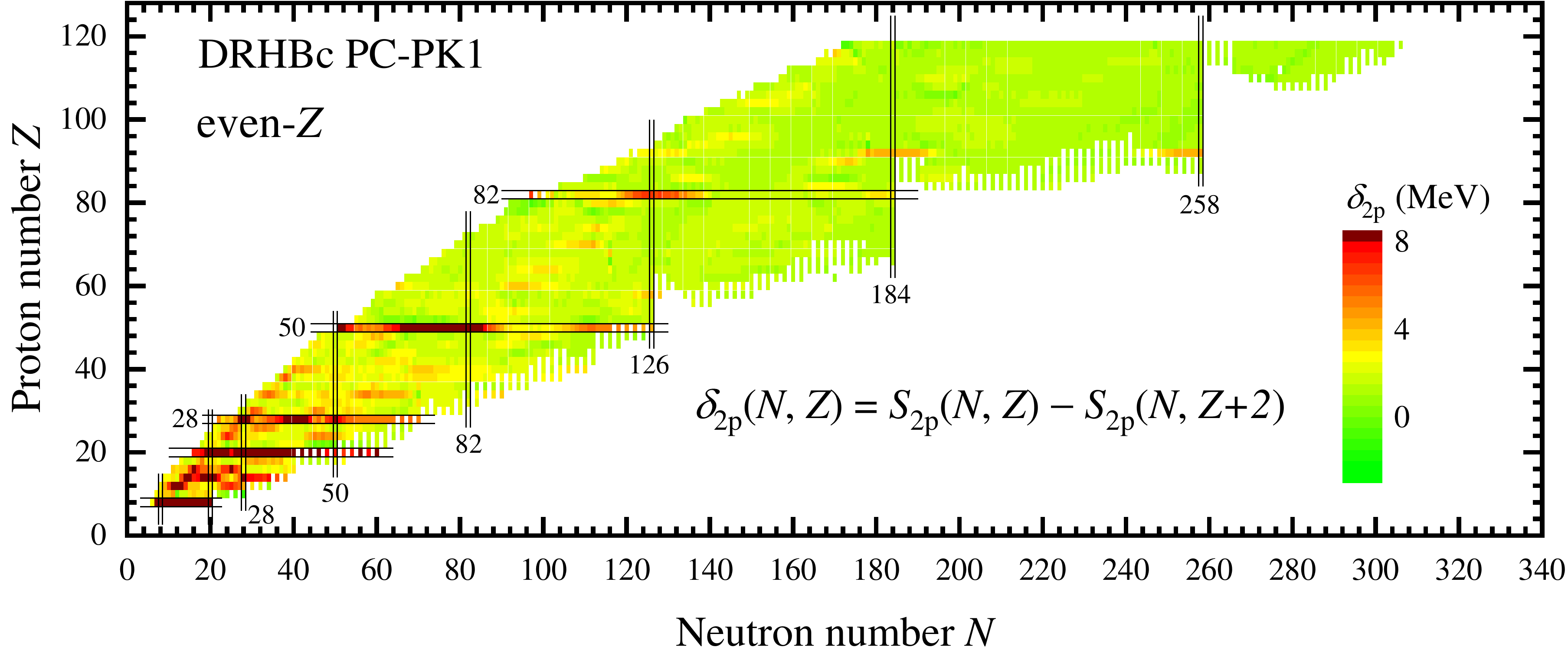}
  \caption{(Color online) Two-proton gaps $\delta_\mathrm{2p}$ of bound even-$Z$ nuclei with $8 \le Z \le 120$ in the DRHBc calculations with PC-PK1 scaled by colors.}
\label{fig9}
\end{figure*}

Figure~\ref{fig9} shows the two-proton gaps $\delta_\mathrm{2p}$ of bound even-$Z$ nuclei with $8 \le Z \le 120$ in the DRHBc calculations with PC-PK1.
$\delta_\mathrm{2p}$ of even-odd nuclei is consistent with their adjacent even-even ones.
In addition, the traditional magic numbers $Z= 8, 20, 28, 50$, and $82$ are obvious from the peaks in Fig.~\ref{fig9}.
Similar to $\delta_\mathrm{2n}$, the peaks of $\delta_\mathrm{2p}$ generally become lower with the increase of proton number, indicating the weakening of the shell effect from light to heavy nuclei.
Besides, in $90 \le N \le 100$ for $Z=50$ and $140 \le N \le 180$ for $Z=82$, where the nuclear shapes obviously deviate from spherical shape, as shown in Fig.~\ref{fig14}, the peaks become weaker or even vanished.
Due to the absence of the calculations for $Z=122$ isotopes, the $\delta_\mathrm{2p}$ for $Z=120$, which was predicted to be the next proton magic number~\cite{Zhang2005NPA,Li2014PLB}, has not been extracted.
Apart from the proton magic numbers, there are some smaller peaks in $\delta_\mathrm{2p}$.
For $Z=14$ isotopes, the even-odd nuclei present pronounced peaks of $\delta_\mathrm{2p}$ in the region of $N\sim 14$, $20$, and $30$, similar to the even-even nuclei in Ref.~\cite{Zhang2022mass}.
As seen in Fig.~\ref{fig14}, most of the even-odd isotopes with $Z=14$ are oblate in their ground states.
This can further support the oblate subshell for light nuclei~\cite{NMBP1980}.
There is a peak at $Z = 92$ near the regions of $N\sim 184$ and $258$, which has been considered as a pseudo shell~\cite{Geng2006CPL}.
In Fig.~\ref{fig8} and Fig.~\ref{fig9}, $\delta_\mathrm{2n}$ and $\delta_\mathrm{2p}$ both display pronounced peaks along the $Z \sim N$ band, which are possibly related to the Wigner energy.
Further investigations on the relation between the peaks of two-nucleon gaps along the $Z \sim N$ band and the Wigner energy would be quite interesting.

\subsection{Alpha decay energies}
\begin{figure*}[htbp]
  \centering
  \includegraphics[width=1.0\textwidth]{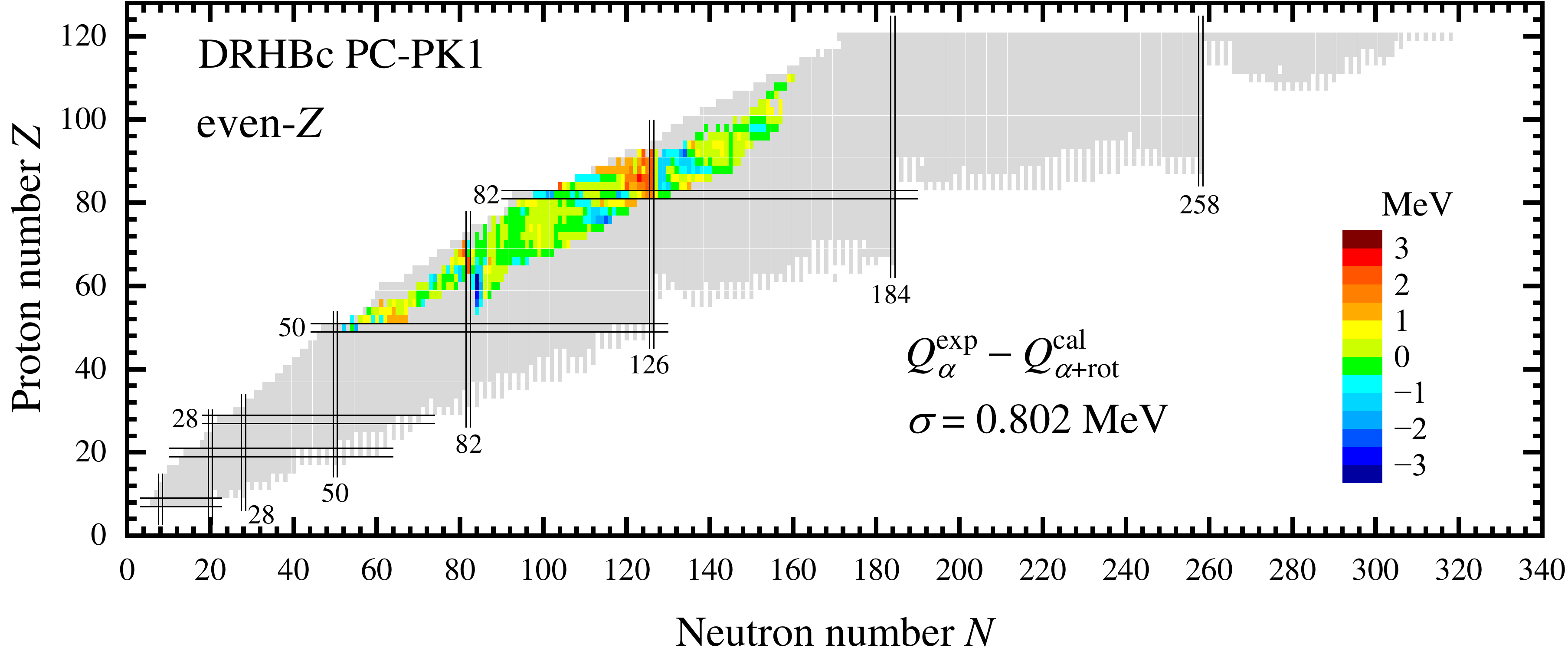}
  \caption{(Color online) $\alpha$ decay energy $Q_\alpha$ differences between the data~\cite{AME2020(3)} and the DRHBc calculations with PC-PK1 for nuclei with $Q_\alpha > 0$, scaled by colors.}
\label{fig10}
\end{figure*}

The $\alpha$ decay energies $Q_\alpha$ can be extracted via the formula,
\begin{equation}
Q_\alpha = E_{\mathrm{b}}(Z-2,N-2)+E_{\mathrm{b}}(2,2)-E_{\mathrm{b}}(Z,N).
\end{equation}
In Fig.~\ref{fig10}, the differences of $Q_\alpha$ between the DRHBc calculations and the data~\cite{AME2020(3)} for nuclei with $Q_\alpha>0$ are scaled by colors.
No obvious odd-even staggering can be found in Fig.~\ref{fig10}, indicating that description of $Q_\alpha$ in even-odd nuclei are consistent with even-even ones.
For the 571 even-$Z$ nuclei, the rms deviation $\sigma=0.802$~MeV, and for the even-even nuclei, $\sigma=0.846$~MeV~\cite{Zhang2022mass}.
After including the deformation degrees of freedom, the description of $Q_\alpha$ has been significantly improved from $\sigma\approx2$~MeV in the RCHB~\cite{Zhang2016CPC}.
In comparison, $\sigma = 0.901$ MeV for even-even nuclei with RHB + DD-PC1 and $\sigma=0.939$~MeV with RHB + DD-ME2~\cite{Agbemava2014PRC}.
In the triaxial relativistic Hartree-Bogoliubov (TRHB) calculations with PC-PK1~\cite{Yang2021PRC}, $\sigma = 0.989$~MeV is reduced to $0.552$~MeV after taking into account the beyond-mean-field correlation energies by using the collective Hamiltonian method.
As discussed in Ref.~\cite{Zhang2022mass}, large deviations in the DRHBc calculations appear near the shell closures $N = 82$ and 126, because the calculations overestimate the binding energies for the nuclei with $N = 82$ and 126 and underestimate their near-spherical neighbors. This fact holds true after including the even-odd nuclei, as shown in Fig.~\ref{fig1}. 
In the future, it would be interesting to further improve the description of $Q_\alpha$ in the DRHBc calculations by including the beyond-mean-field correlation energies using the collective Hamiltonian method~\cite{Sun2022CPC} and systematically estimate the $\alpha$ decay half-lives with the DRHBc theory.

\subsection{Rms radii}
\subsubsection{Charge radii}
\begin{figure*}[htbp]
  \centering
  \includegraphics[width=1.0\textwidth]{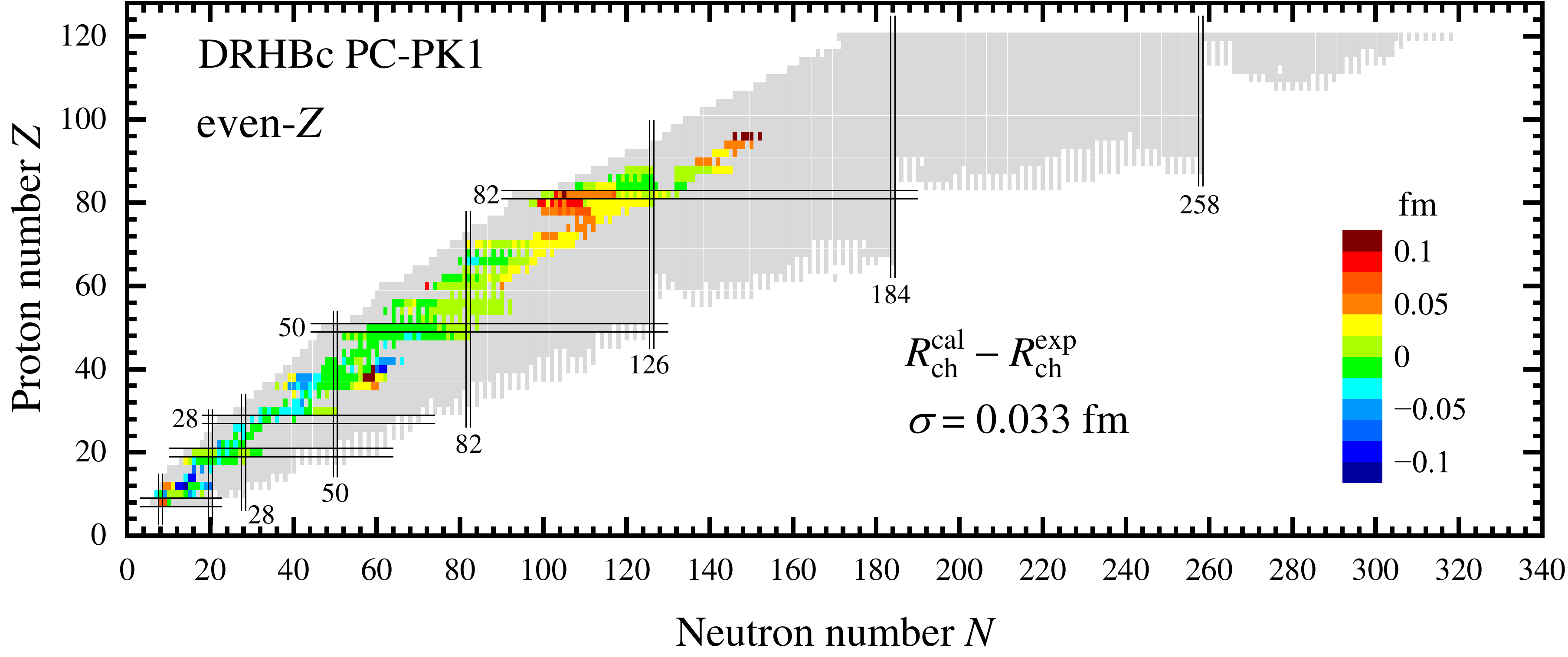}
  \caption{(Color online) For the 620 even-$Z$ nuclei ($8 \le Z \le 120$) with charge radius measured, the deviations between the DRHBc calculations with PC-PK1 and the data are scaled by colors.}
\label{fig11}
\end{figure*}

The nuclear charge radius is the most crucial observable to probe nuclear size as it can be determined through the electromagnetic interaction.
In Fig.~\ref{fig11}, the deviations between the DRHBc calculations and the experimental data are scaled by colors for the 620 even-$Z$ nuclei with charge radius measured~\cite{Angeli2013ADNDT,Li2021ADNDT}.
The DRHBc calculations well reproduce the experimental data and most of the deviations are within $\pm 0.05$~fm.
For the 620 even-$Z$ nuclei, the rms deviation $\sigma=0.033$~fm, and for the 369 even-even nuclei, $\sigma = 0.032$~fm.
As shown in Fig.~\ref{fig11}, the DRHBc theory presents almost the same accuracy in describing charge radii of even-even nuclei and even-odd ones.
After including the deformation degrees of freedom, the description of charge radii has been slightly improved from $\sigma = 0.036$~fm in the RCHB~\cite{Xia2018ADNDT}.
Finally, we note that some large discrepancies mentioned in Ref.~\cite{Zhang2022mass} for even-even nuclei can also be found for the neighboring even-odd nuclei, such as the discrepancies in some light nuclei with $Z <$ 20 and the overestimation in some Pt, Hg, and Cm isotopes.

\subsubsection{Neutron radii}
\begin{figure*}[htbp]
  \centering
  \includegraphics[width=0.7\textwidth]{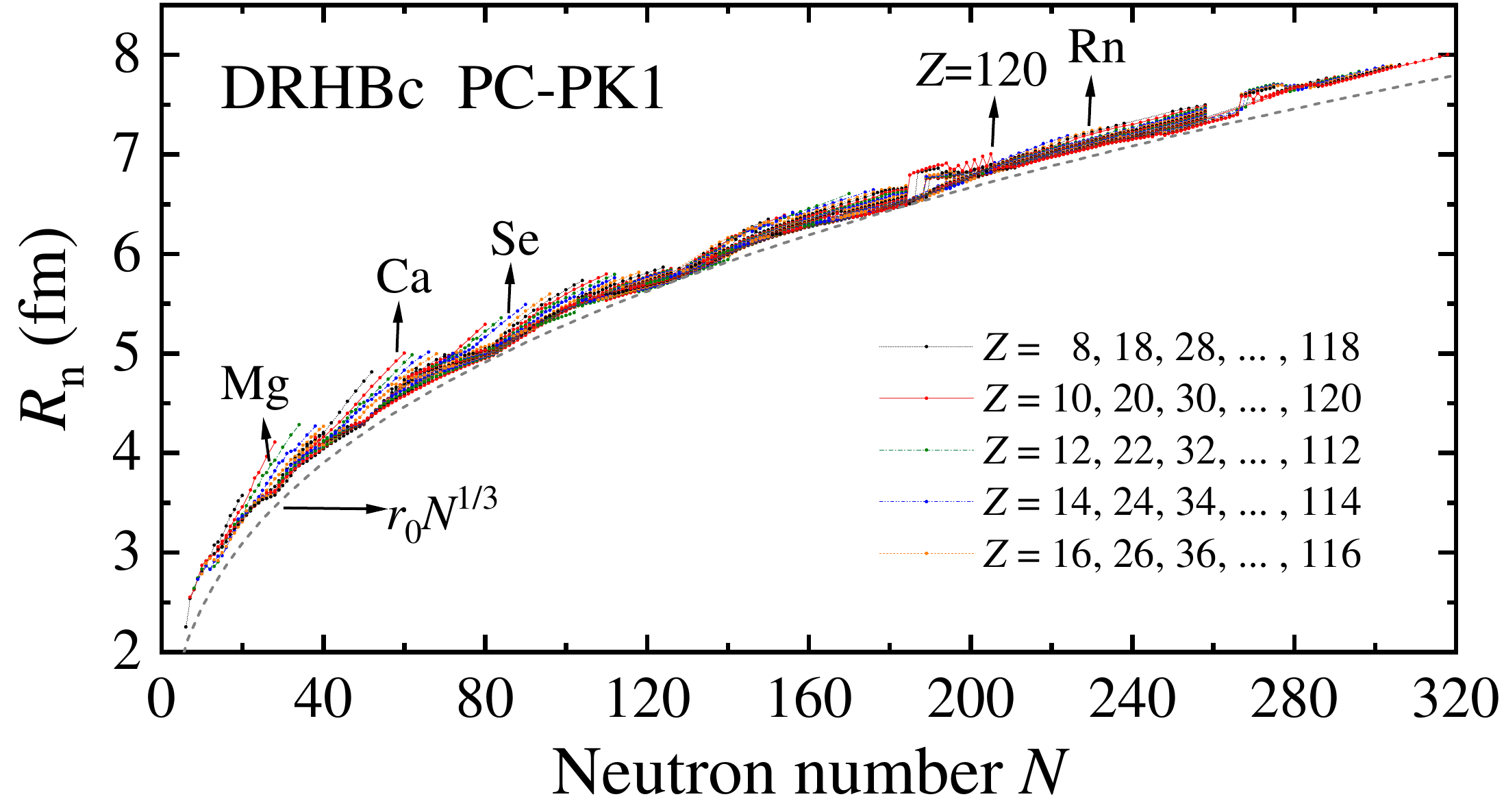}
  \caption{(Color online) Neutron rms radii for even-$Z$ nuclei with $8 \le Z \le 120$ from the DRHBc calculations with PC-PK1 as a function of the neutron number. The empirical formula $r_0 N^{1/3}$ with $r_0 = 1.140$~fm determined from the neutron rms radius of $^{208}$Pb is plotted for guidance.}
\label{fig12}
\end{figure*}

Figure~\ref{fig12} shows the calculated neutron rms radii $R_\text{n}$ for even-$Z$ nuclei, where the empirical formula $R_\text{n} = r_0N^{1/3}$ is drawn as a dashed line for guidance.
After including the even-odd nuclei, the systematic trend of the neutron radii still follows the empirical formula.
Note that pronounced deviations from the empirical formula can be usually found in some extremely neutron-rich nuclei near the drip line, e.g., the near-drip-line Mg and Ca nuclei, and it can be regarded as one of the signals for the halo or giant halo phenomena~\cite{Zhou2010PRC,Li2012PRC,Meng2002PRC,Zhang2003SCG,Terasaki2006PRC}, as already discussed in Ref.~\cite{Zhang2022mass}.
In heavier regions, pronounced deviations from the empirical formula can be also found when the neutron number lies in the middle of two closed shells. Such deviations mainly come from the deformation effect. The abrupt increases in $R_\text{n}$ at $N \sim 190$ and $N \sim 270$ come from the shape changes from a spherical shape respectively around neutron magic numbers $184$ and $258$ to a large prolate shape ($\beta_2 \approx 0.5$). 
A staggering at $N \sim 200$ is due to the fact that neutron radii of even-odd nuclei are larger than their neighboring even-even ones.
These even-odd nuclei have oblate shapes with $\beta_2 \approx -0.43$, whereas their neighboring even-even nuclei have prolate shapes with $\beta_2 \approx 0.38$.

\begin{figure*}[htbp]
  \centering
  \includegraphics[width=1.0\textwidth]{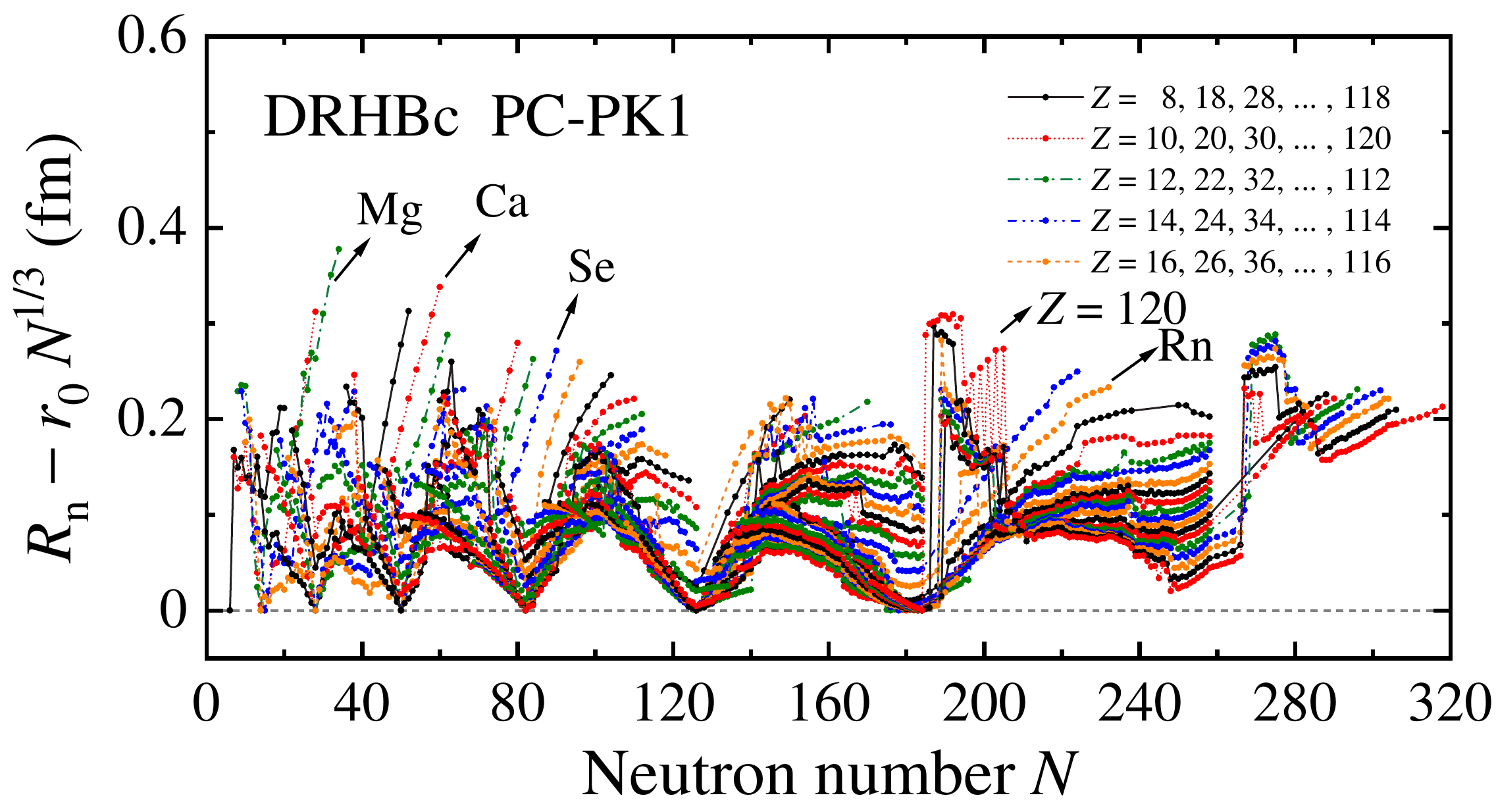}
  \caption{(Color online) Deviations of the DRHBc calculated neutron rms radii from the empirical formula $r_0 N^{1/3}$ for even-$Z$ nuclei with $8 \le Z \le 120$, in which the smallest ratio $R_n/N^{1/3}$ is chosen as $r_0$ for each isotopic chain to ensure non-negative values.}
\label{fig13}
\end{figure*}
 
To have a close inspection, in Fig.~\ref{fig13}, the differences of the calculated neutron rms radii from the empirical formula are depicted, where for each isotopic chain, $r_0$ is chosen as the smallest ratio $R_n/N^{1/3}$.
In comparison with the corresponding plot for even-even nuclei (cf. Fig. 8 in Ref.~\cite{Zhang2022mass}), the evolution of the $R_\text{n}-r_0 N^{1/3}$ value keeps its main characteristics after including the even-odd nuclei.
The nearly vanishing deviation position corresponds to the almost most stable isotope, including the neutron magic numbers $N=20,28,50,82,126$, $184$, and $258$.
Away from the nearly vanishing deviation position, the deviation increases monotonically.
The pronounced deviations near the neutron drip line may indicate the existence of halo or giant halo in some isotopic chains, such as Mg and Ca.
The large deviations at $N\sim 200$ and $270$ correspond to the spherical to strongly prolate shape changes.
The radius staggering at $N\sim 200$ in Fig.~\ref{fig13} yields a more drastic staggering in $R_\text{n}-r_0 N^{1/3}$ for $Z=120$ isotopic chain.
It corresponds to shape changes from prolate in even-even nuclei to oblate in even-odd nuclei.

\subsection{Quadrupole deformation and potential energy curves}
\subsubsection{Quadrupole deformation}\label{qude}
\begin{figure*}[htbp]
  \centering
  \includegraphics[width=1.0\textwidth]{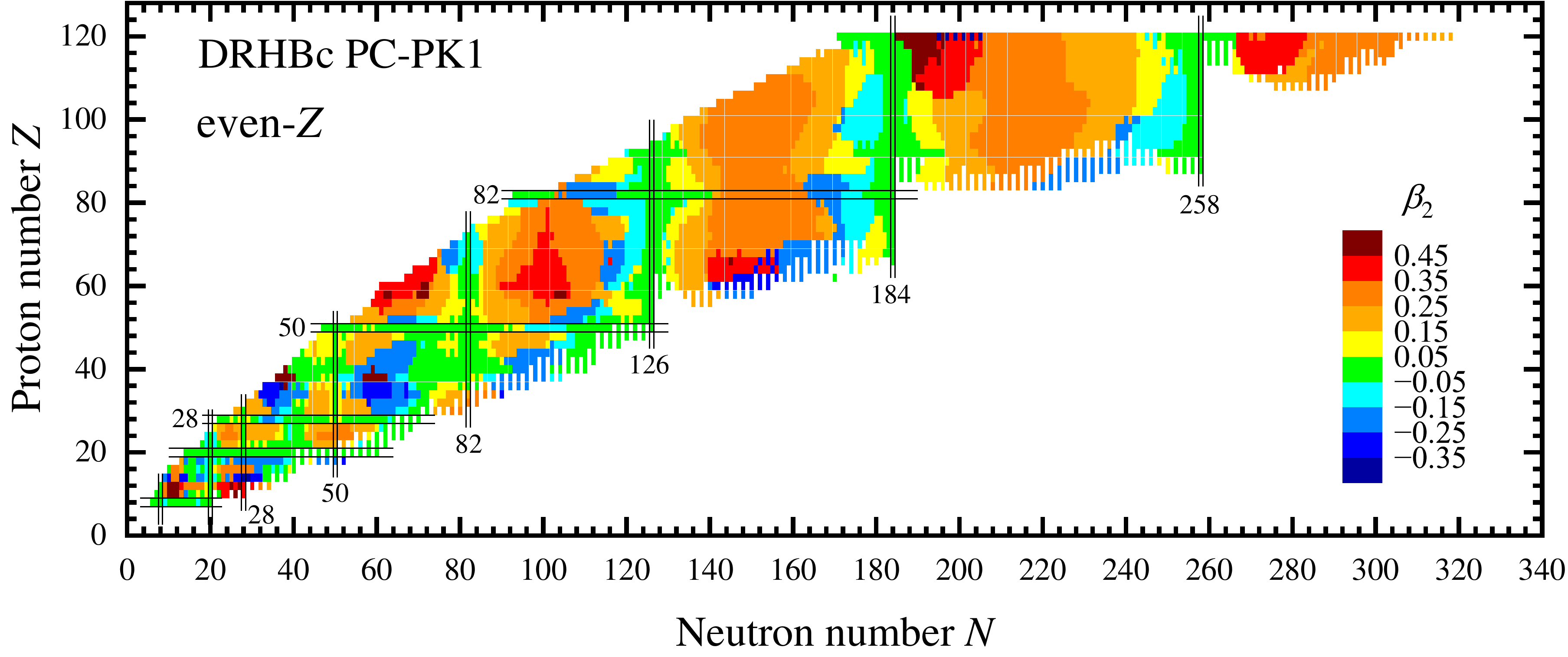}
  \caption{(Color online) Quadrupole deformations $\beta_2$ from the DRHBc calculation with PC-PK1 for bound even-$Z$ nuclei with $8 \le Z \le 120$ scaled by colors.}
\label{fig14}
\end{figure*}

In Fig.~\ref{fig14}, the ground-state quadrupole deformations $\beta_2$ given by the DRHBc calculations are presented for even-$Z$ nuclei with $8 \le Z \le 120$.
There are 539 spherical even-even nuclei and 63 even-odd ones. The significant difference in the number of spherical nuclei indicates the polarization effect of odd nucleon.
The spherical and near-spherical nuclei often appear around neutron and proton magic numbers as expected.
Nuclei near proton or neutron magic number deviate far from spherical shape in some regions, for example, for $N = 28, 50,$ and $82$ with $Z\sim 10, 20$, and $32$, respectively.
These regions are consistent with the weakening or disappearance of two-nucleon gaps in Fig.~\ref{fig8}, showing the weakening or even collapse of the traditional shell closures.

There are 3173 prolate nuclei (1559 even-even and 1614 even-odd ones), and 1054 oblate ones (486 even-even and 568 even-odd ones).
The number of prolate nuclei is much larger than that of oblate ones. The number of even-even prolate (oblate) nuclei is quite close to that of even-odd prolate (oblate) ones.
The well-deformed nuclei often appear in middle of the shell.
Exceptions are around the neutron magic numbers $N=184$ and $N=258$, where nuclei are very strongly deformed.

Compared with the even-even case in Ref.~\cite{Zhang2022mass}, the nuclear shape evolution in an isotopic chain between two closed shells remains the same after including the even-odd nuclei.
Along an isotopic chain, the prolate shapes often develop after a major shell, and the oblate shapes often occur at the end of the major shell.
The dominance of nuclei with prolate shape over those with oblate shape is mainly attributed to more downsloping orbitals on the prolate side~\cite{CASTEN2000,Horoi2010PRC}.
In Ref.~\cite{Guo2023PRC}, taking Te, Xe, and Ba isotopes as examples, the prolate shape dominance was investigated with the DRHBc theory.

In Fig.~\ref{fig14}, there are several noteworthy shape changes.
The prolate-oblate shape changes can be found around $N = 120$ and 160 for $50 \le Z \le 82$.
Some of these shape changes are attributed to the competition between the prolate and oblate minimum (for details cf. Fig.~\ref{fig15}) and some have relations with the triaxial deformation.
For instance, the triaxiality in the light nuclear region with $Z < 20$ and in the heavier region with $114 \le N \le 120$ for $54 \le Z \le 78$ has been investigated~\cite{Yao2010PRC,Yao2011PRC,Yang2021PRC,Zhang2023PRCL}.
There are abrupt shape changes from spherical to large prolate shape after the predicted shell closures $N=184$ and $258$, related to the sudden changes in neutron rms radii shown in Fig.~\ref{fig13}.

From Fig.~\ref{fig14}, the deformation parameters of most even-odd nuclei are consistent with their even-even neighbors.
However, some shape staggering between neighboring even-even and even-odd nuclei can also be observed, such as the staggering between prolate and oblate shape around $N = 160$ for $Z\sim 60$, and $N = 200$ for $Z = 120$.
Such staggering in deformations corresponds to the staggering in neutron rms radii, appearing in Fig.~\ref{fig13}.

\subsubsection{Potential energy curves}
\begin{figure*}[htbp]
  \centering  
  \includegraphics[width=1.0\textwidth]{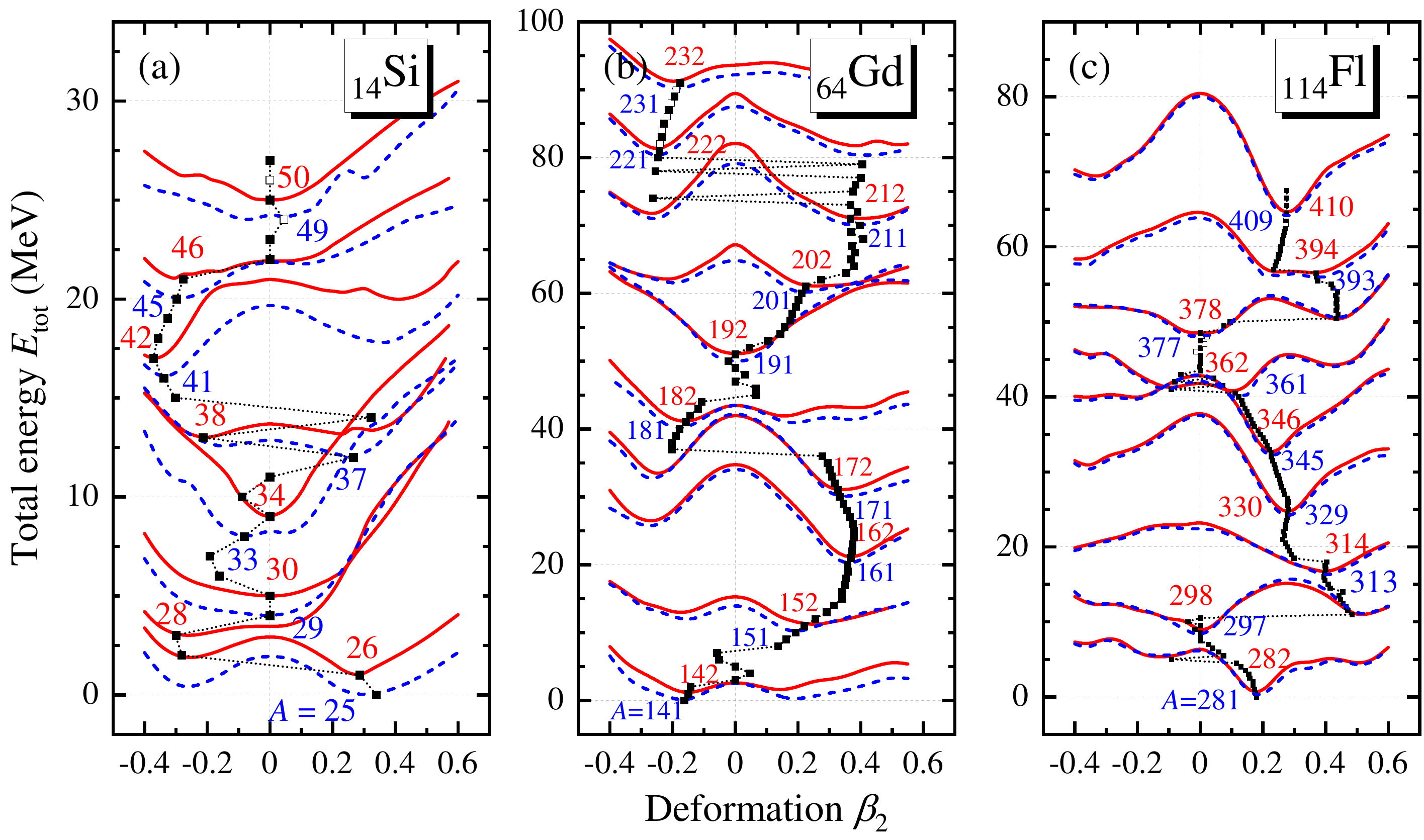}
  \caption{(Color online) The evolution of potential energy curves of $^{25,29,\cdots,49}$Si (a), $^{141,151,\cdots,231}$Gd (b), and $^{281,297,\cdots,409}$Fl (c), together with their neighboring $^{26,30,\cdots,50}$Si (a), $^{142,152,\cdots,232}$Gd (b), and $^{282,298,\cdots,410}$Fl (c) from the constrained DRHBc calculations with PC-PK1. For clarity, in each panel, the PEC of the lightest isotope ($^{25}$Si, $^{141}$Gd, and $^{281}$Fl, respectively) is renormalized to its ground state (filled square), and other PECs are shifted upward one by one by $1$ MeV for Si and Gd and by 0.5 MeV for Fl, per increasing 1 neutron. The ground-state deformations are denoted by squares. For the convenience in comparison with the shell model calculations, the PEC of $^{28}$Si is added.}
\label{fig15}
\end{figure*}

The constrained calculations, where the quadrupole deformation parameter $\beta_2$ is constrained to a given value, is performed to obtain the potential energy curves~\cite{Zhang2020PRC}. In Fig.~\ref{fig15}, the potential energy curves (PECs) of even-odd nuclei $^{25,29,\cdots,49}$Si, $^{141,149,\cdots,229}$Gd, and $^{281,297,\cdots,409}$Fl, together with their neighboring even-even nuclei $^{26,30,\cdots,50}$Si, $^{142,152,\cdots,232}$Gd, and $^{282,298,\cdots,410}$Fl are presented for a better understanding of the shape evolution.
The evolution of ground-state deformation is also depicted for guidance.
The global minima of these PECs obtained from constrained calculations are completely consistent with the ground states from unconstrained ones, verifying the self-consistency of the DRHBc calculations.
The PEC of an even-odd nucleus highly resembles that of its even-even neighbor in most cases, reflecting their similar ground-state deformations.

In the Si isotopic chain, compared with the steep PEC of $^{34}$Si, the PEC of $^{33}$Si is flat with two shallow local minima, which highlights the importance of the polarization effect of the unpaired odd nucleon in a specific light nucleus. 
The PEC of $^{41}$Si is similar with the PEC of $^{42}$Si with a steep minimum in $\beta_2 \approx -0.3$ and a relatively shallow minimum at $\beta_2\approx 0.35$.
The absence of a spherical minimum in the PEC is a natural consequence of the collapse of $N=28$ shell closure in this neutron-rich region.

For heavier isotopic chains, the influence of the polarization effect of the unpaired nucleon is relatively weaker. Therefore, the PECs of Gd and Fl even-odd isotopes are quite similar to those of their even-even neighbors and thus the discussions about Gd and Fl isotopes in Ref.~\cite{Zhang2022mass} still hold.
In the Gd isotopic chain, there are prolate-oblate shape changes around $N = 110$ and $N = 160$. Just as discussed in Ref.~\cite{Zhang2022mass}, the TRHB calculations with PC-PK1~\cite{Yang2021PRC} predict several triaxially deformed nuclei in the former region. In the latter one, both the DRHBc and TRHB calculations predict prolate-oblate shape changes, indicating the possible shape coexistence in nuclei around $N\sim160$ for $Z\sim64$.
In the Fl isotopic chain, there are nearly degenerate spherical and prolate minima in the PECs of $^{297,298}$Fl and $^{377,378}$Fl, further explaining the sudden changes of ground-state deformation from $\beta_2\approx 0$ to $\beta_2 \gtrsim 0.4$ after $N=184$ and $258$, respectively.

\subsection{Neutron density distributions}\label{Ndd}
\begin{figure*}[htbp]
  \centering
  \includegraphics[width=1.0\textwidth]{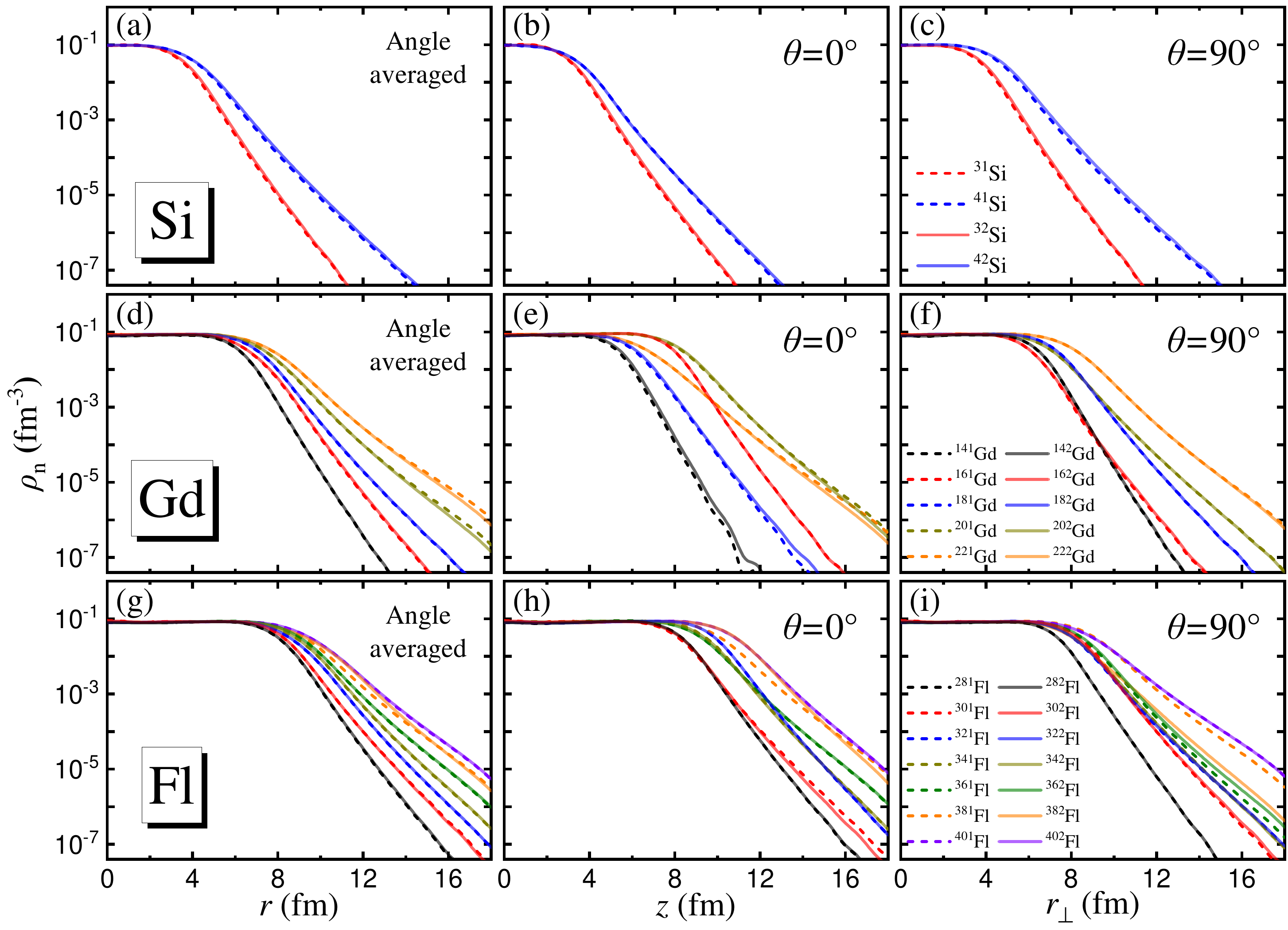}
  \caption{(Color online) Angle averaged neutron density distribution (Angle averaged), the neutron density distribution along the symmetry axis $z$ ($\theta=0^\circ$), and that perpendicular to the symmetry axis with $r_\perp = \sqrt{x^2 + y^2}$ ($\theta=90^\circ$), for selected even-odd isotopes $^{31,41}$Si (a, b, c), $^{141,161,\cdots,221}$Gd (d, e, f), and $^{281,301,\cdots,401}$Fl (g, h, i), together with their neighboring even-even isotopes $^{32,42}$Si (a, b, c), $^{142,162,\cdots,222}$Gd (d, e, f), and $^{282,302,\cdots,402}$Fl (g, h, i) in the DRHBc calculations with PC-PK1.}
\label{fig16}
\end{figure*}

The DRHBc theory can provide an adequate description of halos in deformed nuclei as the continuum, deformation effects, large spatial distributions, and the coupling among them are included in a self-consistent way~\cite{Zhou2010PRC}.
The evolution of neutron density distributions with neutron number can provide information about possible neutron skin or halos.

Figure~\ref{fig16} shows the neutron density profiles of selected even-odd isotopes $^{31,41}$Si, $^{141,161,\cdots,221}$Gd, and $^{281,301,\cdots,401}$Fl, together with their neighboring even-even isotopes $^{32,42}$Si, $^{142,162,\cdots,222}$Gd, and $^{282,302,\cdots,402}$Fl.
The angle averaged neutron density distributions, the neutron density distributions along the symmetry axis ($\theta=0^\circ$), and those perpendicular to the symmetry axis ($\theta=90^\circ$) are depicted in the left, middle, and right panels, respectively.

For the Si isotopes, the angle averaged density distributions in $^{41,42}$Si are similar and more extended than those in $^{31,32}$Si.
In Figs.~\ref{fig16}(b) and \ref{fig16}(c) for $\theta=0^\circ$ and $90^\circ$ respectively, the density distributions in $^{41,42}$Si remain similar and more extended than those in $^{31,32}$Si.
The density distributions at $\theta=90^\circ$ in Fig.~\ref{fig16}(c) are more extended than those at $\theta=0^\circ$ in Fig.~\ref{fig16}(b).
For example, for the density distribution at 10~fm in $^{42}$Si, $\rho_n = 2.28 \times 10^{-6} ~\text{fm}^{-3}$ for $\theta=0^\circ$, while $\rho_n = 1.97 \times 10^{-5} ~\text{fm}^{-3}$ for $\theta=90^\circ$.
This reflects the oblate shape characteristic of these isotopes.

For the Gd isotopes, in Fig.~\ref{fig16}(d), the angle averaged density distributions extend monotonically with neutron number, and the density distributions in even-odd nuclei are similar to those in their even-even neighbors.
In Fig.~\ref{fig16}(e) for $\theta=0^\circ$, the density distributions in even-odd nuclei remain similar to those in their even-even neighbors.
The density distributions in $^{161,162}$Gd are more extended than those in $^{181,182}$Gd, because $^{161,162}$Gd are prolate with $\beta_2 \approx 0.22$ and $^{181,182}$Gd are oblate with $\beta_2 \approx -0.17$.
The density distributions in $^{201,202}$Gd are more extended than those in $^{221,222}$Gd, because $^{201,202}$Gd are prolate with $\beta_2 \approx 0.32$ and $^{221,222}$Gd are oblate with $\beta_2 \approx -0.24$.
In Fig.~\ref{fig16}(f) for $\theta=90^\circ$, the density distributions in even-odd nuclei remain similar to those in their even-even neighbors.
The density distributions extend monotonically with neutron number, except the crossings between $^{141,142}$Gd and $^{161,162}$Gd, and $^{181,182}$Gd and $^{201,202}$Gd at $r_{\perp} \approx 9$ fm.

For the Fl isotopes, in Fig.~\ref{fig16}(g), the angle averaged density distributions extend monotonically with neutron number.
The density distributions in most even-odd nuclei are similar to those in their even-even neighbors, except the difference between $^{381}$Fl and $^{382}$Fl at $10\lesssim r\le14$~fm.
In Fig.~\ref{fig16}(h) for $\theta=0^\circ$, the density distributions in most even-odd nuclei remain similar to those in their even-even neighbors, except the difference between $^{381}$Fl and $^{382}$Fl at $8\lesssim z\le14$~fm.
The density distributions extend monotonically with neutron number.
In the range of $z \ge 14$ fm, the density distributions in $^{321,322}$Fl are close to those in $^{341,342}$Fl, and the density distributions in $^{382}$Fl are close to those in $^{401,402}$Fl.
In Fig.~\ref{fig16}(i) for $\theta=90^\circ$, the density distributions in most even-odd nuclei remain similar to those in their even-even neighbors, except $^{381}$Fl where density distributions are more extended than those in $^{382}$Fl.
This is due to the sudden increase of deformation from $\beta_2 = 0.096$ at $^{381}$Fl to $\beta_2 = 0.434 $ at $^{382}$Fl.
The density distributions extend almost monotonically with neutron number.
The density distributions in $^{301,302}$Fl, $^{321,322}$Fl, $^{341,342}$Fl, $^{361,362}$Fl and $^{382}$Fl are close to each other.
The density distributions in $^{381}$Fl are more extended and close to those in $^{401,402}$Fl due to the deformation effects.
In conclusion, the neutron density distributions manifest not only the neutron diffuseness with the increasing neutron number but also the deformation effects.

\subsection{Neutron potential and diffuseness}
\begin{figure*}[htbp]
  \centering
  \includegraphics[width=1.0\textwidth]{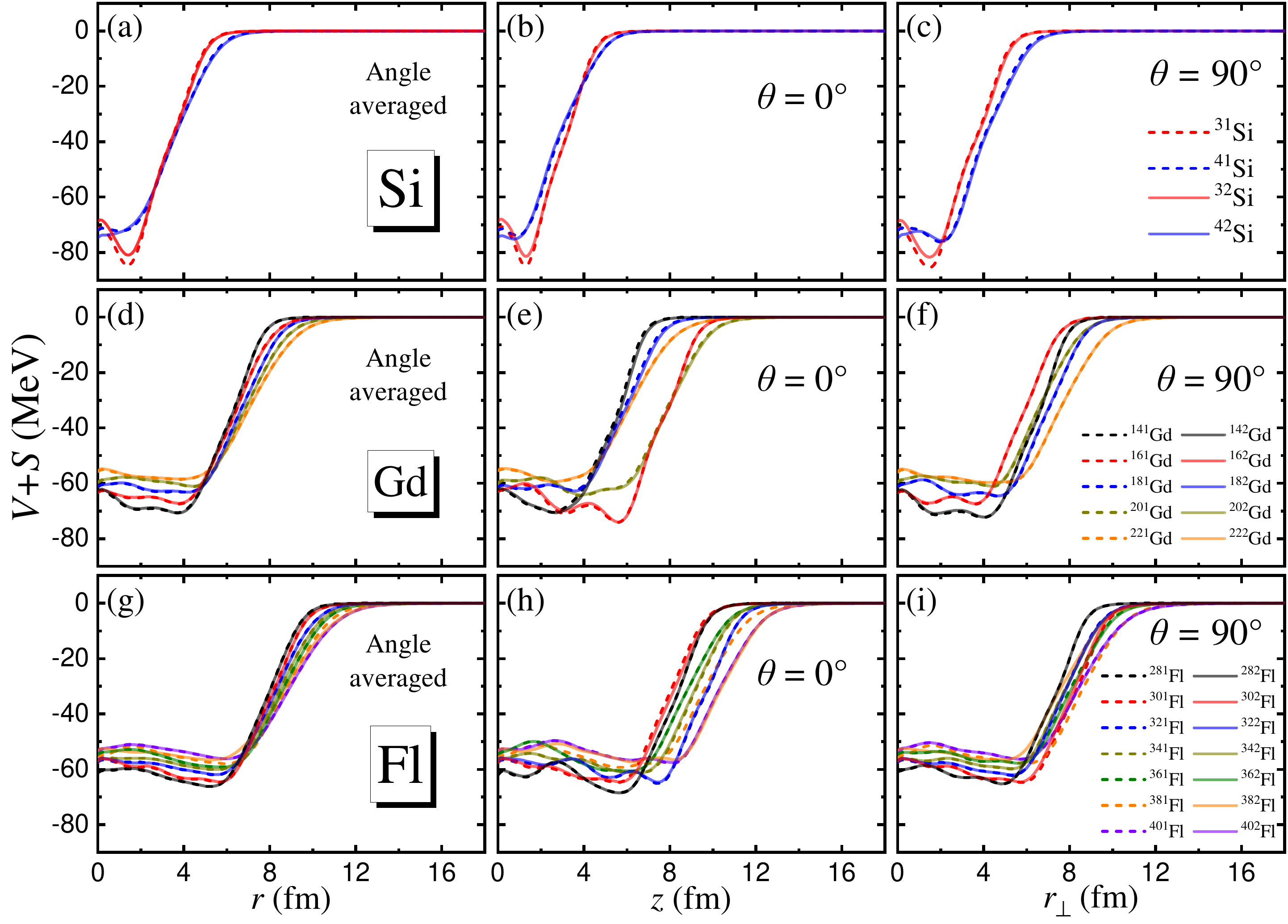}
  \caption{(Color online) Same as Fig.~\ref{fig16}, but for the neutron mean-field potential $V+S$.}
\label{fig17}
\end{figure*}

The mean-field potential, the vector plus scalar potential $V(\bm r) + S(\bm r)$, in DRHBc is calculated self-consistently including the continuum, the deformation, the pairing correlation and the blocking effects.
The examination on the diffuseness of the potential with neutron number can provide valuable guidance for the nuclear models based on mean-field~\cite{Wang2014PLB,Meng1999NPA}.

The neutron mean-field potentials for selected even-odd isotopes $^{31,41}$Si, $^{141,161,\cdots,221}$Gd, and $^{281,301,\cdots,401}$Fl, together with their neighboring even-even isotopes $^{32,42}$Si, $^{142,162,\cdots,222}$Gd, and $^{282,302,\cdots,402}$Fl, are presented in Fig.~\ref{fig17}, in terms of the angle averaged potential and those along ($\theta=0^\circ$) and perpendicular to ($\theta=90^\circ$) the symmetry axis.

For the Si isotopes, in Fig.~\ref{fig17}(a), the angle averaged potentials in $^{31}$Si and $^{41}$Si are similar to those in $^{32}$Si and $^{42}$Si, respectively, except the slight difference in the central depletion in the potential for $^{31,32}$Si.
In $^{41,42}$Si, the depth of the potentials decreases, and the surface of the potentials moves outward in comparison with $^{31,32}$Si.
In Fig.~\ref{fig17}(b) for $\theta=0^\circ$ and Fig.~\ref{fig17}(c) for $\theta=90^\circ$, the potentials in $^{31}$Si and $^{41}$Si are respectively similar to those in $^{32}$Si and $^{42}$Si.
The potentials in $^{41,42}$Si at $\theta=90^\circ$ are more diffused than those at $\theta=0^\circ$, due to the oblate deformation $\beta_2 < -0.3$ in $^{41,42}$Si.
Similar to $^{32}$Si in Ref.~\cite{Zhang2022mass}, the central depletion in the neutron potential for $^{31}$Si is related with the loss of 2$s$ components due to the deformation effects.

For the Gd isotopes, in Fig.~\ref{fig17}(d) for the angle averaged potential, the depth decreases monotonically, the surface moves outward, and the diffuseness increases with neutron number.
The potentials in even-odd nuclei are similar to those in their even-even neighbors.
In Fig.~\ref{fig17}(e) for $\theta=0^\circ$, the potentials in even-odd nuclei remain similar to those in their even-even neighbors.
The most diffused potentials are those in $^{161,162}$Gd and $^{201,202}$Gd.
This is because $^{161,162}$Gd and $^{201,202}$Gd are prolate and the others are oblate.
In Fig.~\ref{fig17}(f) for $\theta=90^\circ$, the potentials in even-odd nuclei remain similar to those in their even-even neighbors.
The depth of the potentials decreases monotonically with neutron number.
The most diffused potentials are those in $^{221,222}$Gd.
Due to their prolate deformation, the potentials in $^{161,162}$Gd are less diffused than those in $^{141,142}$Gd.

For the Fl isotopes, in Fig.~\ref{fig17}(g) for the angle averaged potential, the depth decreases monotonically, the surface moves outward, and the diffuseness increases with neutron number.
The potentials in most even-odd nuclei are similar to those in their even-even neighbors, except the difference between $^{381}$Fl and $^{382}$Fl at $5\lesssim r\le8$~fm and $10\lesssim r\le12$~fm.
In Fig.~\ref{fig17}(h) for $\theta=0^\circ$, the potentials in most even-odd nuclei remain similar to those in their even-even neighbors, except $^{381}$Fl where the depth of the potential increases and the potential is less diffused than that in $^{382}$Fl.
The depth of the potential decreases monotonically and the difuseness increases with neutron number in general.
The potential in $^{382}$Fl is as diffused as those in $^{401,402}$Fl and much more diffused than that in $^{381}$Fl.
The potential in $^{301,302}$Fl is less diffused than $^{281,282}$Fl.
In Fig.~\ref{fig17}(i) for $\theta=90^\circ$, the potentials in most even-odd nuclei remain similar to those in their even-even neighbors, except the difference between $^{381}$Fl and $^{382}$Fl at $r_{\perp} \ge 5$ fm.
The depth of the potential decreases monotonically and the diffuseness increases with neutron number in general.
The potential in $^{381}$Fl is as diffused as those in $^{401,402}$Fl and much more diffused than that in $^{382}$Fl.
The differences in the potential between $^{381}$Fl and $^{382}$Fl are due to the sudden increase of deformation from $^{381}$Fl to $^{382}$Fl, as discussed in Sec.\ref{Ndd}.


\section{Summary}\label{summary}

In summary, we have performed systematic studies of all even-$Z$ nuclei with $8 \le Z \le 120$ from the proton drip line to the neutron drip line by using the DRHBc theory with the density functional PC-PK1, extending the work in Ref.~\cite{Zhang2022mass} to even-odd nuclei.

There are 4829 even-$Z$ nuclei with $8 \le Z \le 120$ in total predicted to be bound.
The rms deviation from the experimental binding energies of 1244 even-$Z$ nuclei is $1.433$ MeV, which provides one of the most accurate microscopic descriptions for nuclear masses.
The calculated binding energies, two-neutron, two-proton, and one-neutron separation energies, rms radii of neutron, proton, matter, and charge distributions, quadrupole deformations, and neutron and proton Fermi surfaces are tabulated.

The evolution of the nucleon separation energies and their drastic changes at magic numbers are discussed. The traditional magic numbers are reproduced well and new magic numbers $N=184$ and $258$ are predicted.
It is noted that the $S_\mathrm{2n}$ values of the even-odd nuclei with one neutron more than magic numbers are close to the average of their neighboring even-even nuclei.
The odd-even effects of the nucleon separation energies are presented and discussed.

The predicted nuclear masses, nucleon separation energies, and limits of the nuclear landscape are compared with other relativistic and non-relativistic density functional calculations.
The proton drip line predicted is close to those by other models, and generally consistent with experiments.
The neutron drip line predicted is generally more extended in comparison with other models, mainly due to the proper treatment of the continuum and deformation in the DRHBc theory and the adopted density functional.
A peninsula of stability in the superheavy region beyond the neutron drip line in RCHB is predicted in the DRHBc mass table after the inclusion of the deformation.
There are bound even-odd nuclei inside the peninsula of stability in the superheavy region.
Some smaller peninsulas~\cite{Zhang2022mass,Pan2021PRC} consisting of only even-even nuclei are found.

The odd-even mass differences $\Delta_\mathrm{n}$ decrease with the mass number $A$ in general.
The obtained rms deviation from experimental data for 1123 even-$Z$ nuclei is $\sigma = 0.580$~MeV.
The pairing energies generally approach zero near magic numbers while they are maximal near the middle of the shells.
From the vanishing pairing energies, new magic numbers $N=184$ and $258$, $Z=120$, and some subshells are predicted.
From the pronounced pairing energies, the collapse of traditional shell closures near the neutron drip lines at $N = 28$ for $Z \sim 10$, $N = 50$ for $Z \sim 20$, and $N = 82$ for $Z \sim 32$ is shown.

From the two-nucleon gaps, the traditional magic numbers are well reproduced and new magic numbers at $N=184$ and $258$ are predicted.
The vanishing of two-nucleon gaps at $N = 28$, 50, and 82 near the neutron drip lines, $Z = 50$ for $90 \le N \le 100$, and $Z = 82$ for $140 \le N \le 180$ represents the collapse of shell closures and the weakening of the shell effect.
Some peaks at $N=40$ for $Z\sim 20$, $N=162$ for $Z \sim 110$, and $Z=14$ for $N\sim 20$ and 30, can be regarded as signals of possible spherical or deformed subshells.

From the $\alpha$ decay energies extracted, the rms deviation from 571 available experimental data is $\sigma = 0.802$~MeV, which is significantly improved from $\sigma \approx 2$ MeV in the RCHB results~\cite{Xia2018ADNDT}. 
It is interesting to further improve the description by including beyond-mean-field correlations, e.g., by the collective Hamiltonian method.

For the 620 even-$Z$ nuclei with charge radius measured, the obtained rms deviation is $\sigma = 0.033$ fm.
The systematic trend of the neutron rms radii for even-$Z$ nuclei with $8 \le Z \le 120$ generally follows the empirical formula.
There are a few exceptions.
For example, the drastic changes at $N \sim 190$ and 270, and the staggering at $N \sim 200$ are related to the deformation.
The pronounced deviations for some extremely neutron-rich nuclei are possible signals for the halo or giant halo phenomena~\cite{Zhou2010PRC,Li2012PRC,Meng2002PRC,Zhang2003SCG,Terasaki2006PRC}.

From the quadrupole deformations, 602 spherical nuclei including 63 even-odd ones, 3173 prolate nuclei, and 1054 oblate nuclei are identified.
The weakening or even collapse of the traditional shell closures is further demonstrated by the deformation deviating from the spherical shape.
The drastic shape changes on the nuclear landscape are due to the competition between the prolate and oblate minima or the triaxial deformation.
The shape evolution in each isotopic chain can be understood from PECs in constrained calculations.
The PEC of an even-odd nucleus is similar to that of its adjacent even-even ones, with a few exceptions due to the polarization effect of the unpaired nucleon.

Finally, the angle averaged neutron density distributions and neutron mean-field potentials as well as those along and perpendicular to the symmetry axis for selected even-odd Si, Gd, and Fl isotopes and their even-even neighbors are presented.
The neutron density distributions and mean-field potentials in neighboring even-even and even-odd nuclei are quite similar with a few exceptions related to drastic shape changes.
The neutron density distributions manifest the diffuseness with the increase of neutron number and the deformation effects.
For the mean-field potential, in general, the depth decreases monotonically, the surface moves outward, and the diffuseness increases with neutron number.

With the continuum, the deformation, the pairing correlation and the blocking effects self-consistently included, the successful exploration of the DRHBc theory in the nuclear chart has been extended to even-odd nuclei, and the corresponding mass table for even-$Z$ nuclei has been constructed.
Systematic calculations for odd-$Z$ nuclei and the construction of the mass table are in progress.

\section*{Declaration of competing interest}
The authors declare that they have no known competing financial interests or personal relationships that could have appeared to influence the work reported in this paper.

\section*{Data availability}
Data will be made available on on request or be provided as downloadable data files at \url{https://drhbctable.jcnp.org}.

\begin{acknowledgments}

This work was partly supported by the National Natural Science Foundation of China (Grants No.~11935003, No.~11975031, No.~11775099, No.~12141501, No.~12070131001, No.~12075085, No. 12375118, No.~12047568, No.~12205308, No.~12222511, No.~12265012, No.~12275082, No.~12375119, No.~12375119, No.~11975041, No.~12047503, No.~11975237, No.~12305125, No.~U2230207, No.~U2030209, and No.~11961141004), the National Key R\&D Program of China (Contracts No.~2017YFE0116700, No.~2018YFA0404400, No.~2020YFA0406001, No.~2020YFA0406002, and No.~2021YFA1601500), the State Key Laboratory of Nuclear Physics and Technology, Peking University (Grants No.~NPT2023ZX03 and No.~NPT2023KFY02), the Strategic Priority Research Program of Chinese Academy of Sciences (Grants No.~XDB34010000), Natural Science Foundation of Sichuan Province (24NSFSC5910), Guangdong Basic and Applied Basic Research Foundation (2023A1515010936), the High-performance Computing Platform of Peking University, the High-performance Computing Platform of Anhui University, the High Performance Computing Resources in the Institute for Basic Science Research Solution Center, the National Research Foundation of Korea (NRF) funded by Ministry of Science and ICT (2013M7A1A1075764).
This work was partly supported by the National Research Foundation of Korea (Grant Nos.~NRF-2018R1D1A1B05048026, NRF-2020R1A2C3006177, NRF-2020K1A3A7A09080134, NRF-2021R1F1A1060066, NRF-2021R1A6A1A03043957, and NRF-2023R1A2C1005398), the Institute for Basic Science (Grant IBS-R301-D1), and the National Supercomputing Center with supercomputing resources including technical support (KSC-2020-CRE-0329, KSC-2021-CRE-0126, KSC-2021-CRE-0272, KSC-2022-CRE-0333, and KSC-2023-CRE-0170).
YBC and CHL were supported by  National Research Foundation of Korea (NRF) grants funded by the Korea government (Ministry of Science and ICT and Ministry of Education) (No.~2016R1A5A1013277 and No.~2018R1D1A1B07048599).
The results from ITP side are obtained on the High-performance Computing Cluster of ITP-CAS and the ScGrid of the Supercomputing Center, Computer Network Information Center of Chinese Academy of Sciences.
\end{acknowledgments}

\newpage
\section*{Explanation of Tables}

\textbf{Table II. Ground-state properties of even-$Z$ nuclei calculated by the DRHBc theory}

\begin{tabular}{@{}p{0.0in}p{1.5in}p{4.5in}@{}}
& $Z$ & Proton number\\
& $N$ & Neutron number\\
& $A$ & Mass number\\
& $E^{\mathrm{cal}}_{\mathrm{b}}$ & Binding energy from DRHBc calculations\\
& $E^{\mathrm{cal}}_{\mathrm{b}+\mathrm{rot}}$ & Binding energy plus rotational correction energy from DRHBc calculations, which is suggested to be compared with the experimental value\\
& $E^{\mathrm{exp}}_{\mathrm{b}}$ & Binding energy from experimental data\\
& $S_\mathrm{2n}$ & Two-neutron separation energy\\
& $S_\mathrm{2p}$ & Two-proton separation energy\\
& $S_\mathrm{n}$ & One-neutron separation energy\\
& $R_\mathrm{n}$ & Neutron root-mean-square radius \\
& $R_\mathrm{p}$ & Proton root-mean-square radius \\
& $R_{m}$ & Matter root-mean-square radius \\
& $R^{\mathrm{cal}}_{\mathrm{ch}}$ & Charge radius from DRHBc calculations \\
& $R^{\mathrm{exp}}_{\mathrm{ch}}$ & Charge radius from experimental data\\
& $\beta_\mathrm{2n}$ & Neutron quadrupole deformation \\
& $\beta_\mathrm{2p}$ & Proton quadrupole deformation \\
& $\beta_{2}$ & Matter quadrupole deformation \\
& $\lambda_\mathrm{n}$ & Neutron Fermi surface\\
& $\lambda_\mathrm{p}$ & Proton Fermi surface\\
& $m^\pi(N)$ & Quantum numbers $m^\pi$ of the blocked neutron orbital \\
& $\sigma$ & Rms deviations for binding energies and charge radii for each isotopic chain \\
\\

\end{tabular}

\textbf{Note: \\
*: Since PC-PK1 is a non-linear density functional, it encounters high density instability in $^{24,25,26}$Mg and $^{26,28}$Si, similar to that in $^{12}$C by NL1~\cite{Reinhard1988ZPA}.\\
$\dag$: Some nuclei have positive $S_\mathrm{2n}$, $S_\mathrm{2p}$, and $S_\mathrm{n}$ as well as negative $\lambda_\mathrm{n}$ and $\lambda_\mathrm{p}$, but are unbound against multi-nucleon emission.\\
If the pairing energy vanishes, the Fermi energy is chosen to be the energy of the last occupied single-particle state.}

\newpage
\begin{landscape}
\pagestyle{empty}

\newpage
\end{landscape}
\end{CJK*}
\end{document}